\documentstyle[prl,osa,multicol,epsfig]{revtex}

\newcommand{\Lbox}{\mbox{$S_{B}$}}
\newcommand{\freqB}{\mbox{$\omega_{B}$}}

\newcommand{\Lav}{\mbox{$\langle L \rangle$}}

\newcommand{\Lmax}{\mbox{$L_{max}$}}
\newcommand{\grow}{\mbox{$\alpha$}}
\newcommand{\gammaeff}{\mbox{$\gamma_e$}}
\newcommand{\phistar}{\mbox{$\phi^{*}$}}
\newcommand{\Lstar}{\mbox{$L^{*}$}}
\newcommand{\sstar}{\mbox{$s^{*}$}}
\newcommand{\cN}{\mbox{$p$}}

\newcommand{\NB}{\mbox{$l$}}
\newcommand{\xiL}{\mbox{$\Xi$}}
\newcommand{\Mav}{\mbox{$\langle M \rangle$}}

\newcommand{\CV}{\mbox{$c_v$}}

\newcommand{\bb}{\mbox{$\left< b^2 \right>$}}
\newcommand{\Rend}{\mbox{$R_e$}}
\newcommand{\Rgyr}{\mbox{$R_g$}}

\newcommand{\Ree}{\mbox{$\left< R_e^2 \right>$}}
\newcommand{\Rgg}{\mbox{$\left< R_g^2 \right>$}}
\newcommand{\Rs}{\mbox{$R_0$}}

\newcommand{\ks}{\mbox{$k_s$}}
\newcommand{\kr}{\mbox{$k_r$}}
\begin{document}
\sloppy

\title{Dynamical Monte Carlo Study of Equilibrium Polymers : Static Properties}

\author{J.~P.~Wittmer$^1$\thanks{to whom correspondence should be addressed.},
A.~Milchev$^2$, M.~E.~Cates$^1$}

\address{
$^1$Department of Physics and Astronomy, University of Edinburgh,\\
JCMB King's Buildings, Mayfield Road, Edinburgh EH9 3JZ, UK\\
\vskip 0.3truecm
$^2$Institute for Physical Chemistry, Bulgarian Academy of Science,
1113 Sofia, Bulgaria}

\maketitle

\centerline{PACS numbers :
61.25.Hq, 05.40+j, 05.50.+q, 64.60.Cn, 82.35.+t}

\vskip 0.3truecm

\centerline{\today}

\vskip 0.3truecm

\begin{abstract}
We report results of extensive Dynamical Monte Carlo investigations on
self-assembled Equilibrium Polymers (EP) without loops in good solvent.
(This is thought to provide a good model of giant surfactant micelles.)
Using a novel algorithm we are able to
describe efficiently both static and dynamic properties
of systems in which the mean chain length $\Lav$ is effectively
comparable to that of laboratory experiments
(up to 5000 monomers, even at high polymer densities).
We sample up to scission energies of $E/k_BT=15$
over nearly three orders of magnitude in monomer density $\phi$,
and present a detailed crossover study ranging from swollen EP chains
in the dilute regime up to dense molten systems.
Confirming recent theoretical predictions, the mean-chain length is found to
scale as $\Lav \propto \phi^{\grow} \exp(\delta E)$
where the exponents approach $\grow_d=\delta_d=1/(1+\gamma) \approx 0.46$
and $\grow_s = 1/2 [1+(\gamma-1)/(\nu d -1)] \approx 0.6, \delta_s=1/2$
in the dilute and semidilute limits respectively.
The chain length distribution is qualitatively well described
in the dilute limit by the Schulz-Zimm distribution
$\cN(s)\approx s^{\gamma-1} \exp(-s)$ where the scaling variable
is $s=\gamma L/\Lav$.
The very large size of these simulations allows also an accurate
determination of the self-avoiding walk susceptibility exponent
$\gamma \approx 1.165 \pm 0.01$.
As chains overlap they enter the semidilute regime where
the distribution becomes a pure exponential $\cN(s)=\exp(-s)$ with the
scaling variable now $s=L/\Lav$.
In addition to the above results we measure the specific heat per monomer
$\CV$.
We show that the average size of the micelles, as measured by the
end-to-end distance and the radius of gyration, follows a crossover scaling
that is, within numerical accuracy, identical to that of conventional
monodisperse quenched polymers.
Finite-size effects are discussed in detail.
\end{abstract}


\begin{multicols}{2}
\section{Introduction.}
\label{sec:Introction}

\begin{figure}[t]
\centerline{\epsfysize=6cm
\epsfig{file=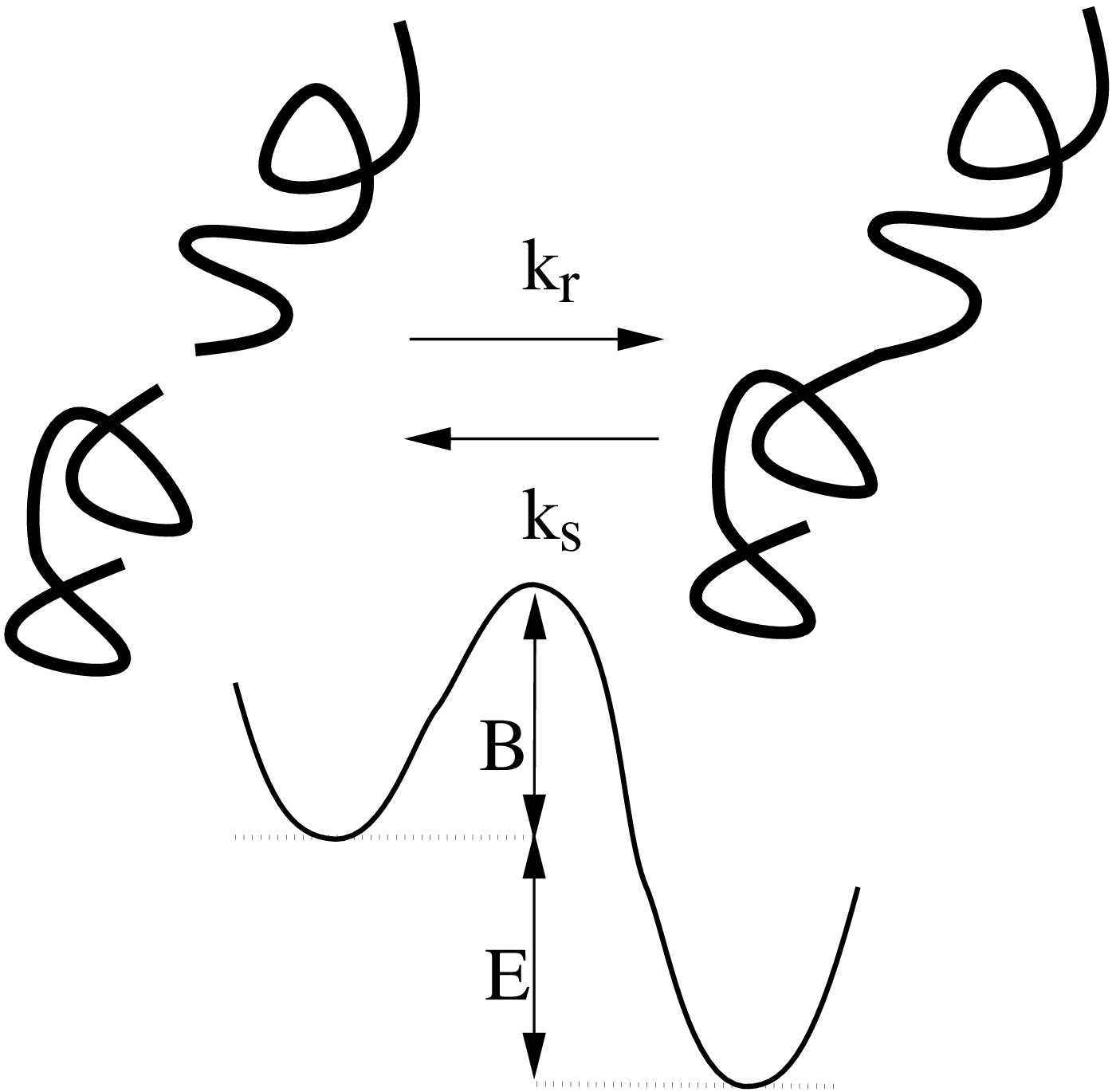,width=50mm,height=40mm,angle=0}}
\vspace*{0.4cm}
{\small FIG.~1. 
Sketch of model:
Bonds of EP chains break and recombine constantly with
rates $\ks=\exp(-(E+B)/k_BT)$ and $\kr=\exp(-B/k_BT)$
depending on the scission energy $E$
and the activation energy barrier $B$,
both supposed to be independent of monomer position and density.
Closed rings and branching of chains are not allowed within our model.
\label{fig:sketchEP}}
\end{figure}

Systems in which polymerization takes place under
condition of chemical equilibrium between the polymers and their
respective monomers are termed ``equilibrium polymers'' (EP).
An important example is that of surfactant molecules forming
long flexible cylindrical aggregates, so-called  {\em giant micelles}
(GM)\cite{CatesCandau}, which break and recombine constantly at random
points along the sequence (see Fig.~\ref{fig:sketchEP}). Similar systems of
EP are formed by liquid sulfur\cite{Sulfur,Pfeuty}, selenium\cite{Selen} and
some protein filaments\cite{Proteins}.
In the surfactant literature (e.g. \cite{CatesCandau}) giant micelles are
often referred to as ``living polymers" although this is potentially
confusing since they are distinct from systems that reversibly polymerize
stepwise, in the presence of fixed number of initiators, for which this term
has previously been reserved \cite{Szwarc56,Greer96}.
As direct imaging methods clearly demonstrate\cite{Clausen} GM, which behave
very much like conventional polymer chains\cite{likepolymer}, may become very
long indeed with contour lengths up to $\approx 1 \mu m$.
However, the constant reversible scission of the chains offers an
additional  stress relaxation mechanism in comparison with conventional
``quenched" polymers whose identity is fixed for all time
\cite{CatesCandau,Cates88,Faet,Bouchaud,OShaug}.

EP are intrinsically polydisperse and their molecular weight distribution
(MWD) in equilibrium is expected \cite{Flory53,CatesCandau} to follow an
exponential decay with chain length. So far, we are not aware of any {\em
direct} experimental measurements of the MWD in  such systems.
Of central interest is the mean chain length, and for GM there has been
experimental and theoretical controversy
\cite{CatesCandau,Schaefer,Schoot,Berret,Schurtenberger}
concerning its dependence on volume fraction,
described by the growth exponent
$\Lav \propto \phi^{\alpha}$.
A scaling theory (summarized below) gives $\alpha \simeq 0.6$; although
this is consistent with some data on ionic micelles at intermediate or
high salt levels \cite{CatesCandau}, a much larger exponent 
$\alpha \simeq 1.2$ is suggested by experiments on lecithin-in-oil reverse micelles and
some nonionic aqueous surfactants \cite{Schurtenberger}. Thus the
experimental evidence concerning the equilibrium growth law of GM remains
controversial.

Given the shortcomings of any approximate analytical treatment
and the difficulties with the laboratory measurements,
numerical experiments, being exact within the
framework of the respective model and able to account explicitly for
various factors that influence experiments, should help much in understanding
the thermodynamic behavior and the properties, both static and dynamic, of EP.
However, up to now only a small number of simulational studies
\cite{MilchevPotts,YA1,YA2,YA3,Kro} exist, in contrast to numerical
experiments with conventional polymers.
Indeed, while the connectivity of polymer chains and the resulting slow
dynamics render computer simulations a demanding task in its own terms,
the scission-recombination processes, which are constantly underway
in EP, impose additional problems for computational algorithms, mainly in
terms of data organization and storage. Since chains constantly break while
other fragments unite into new chains, objects can lose their identity or
gain new ones at each step of the simulation.

In earlier Monte Carlo (MC) simulations on EP\cite{MilchevPotts}
the systems of polydisperse polymer chains were mapped onto an asymmetric
Potts model, in which different spin values were taken to represent bonded
and nonbonded monomers as well as vacancies in a lattice. Such models are
very efficient for studying static properties of EP since at each update of
the lattice all sites are assigned new spin values subject to a Boltzmann
probability; dynamically however this violates the {\em connectivity} of the
chains. Accordingly this approach faithfully reproduces static properties,
but since the kinetics of such models is fictitious these cannot be extended
to study dynamics which is one of our goals.

In a recent work by Y.~Rouault and one of us, a
Dynamical Monte Carlo algorithm (DMC) was proposed\cite{YA1},
based on the highly efficient bond fluctuation model
(BFM)\cite{BFM,Wolferl,Marcus,WPB1} which is known to be very accurate in
reproducing both static and dynamic properties (Rouse behavior)
of polymer chains in  melts and solutions\cite{Wolferl}. However,
the data structure used was based on a quenched polymer algorithm
and was therefore rather slow and memory consuming;
a radical new approach is required. Below we present a
method to deal with these problems efficiently and are thereby able to study
much larger systems.

In several EP systems, the behavior is strongly affected by the presence
of polymeric rings\cite{ringeffect}. For reasons that are not entirely
clear however, ring-formation seem to be negligible in other cases,
included that of GM.
(This does not necessarily exclude a small number of closed loops in GM
systems as may sometimes be seen using directimaging methods\cite{Clausen}.)
Since the latter are among the most widely studied examples of EP, our
results on the ring-free case are presented here. (This elaborates a brief
previous discussion\cite{WMC1}.)  Corresponding results on EP systems
containing rings will be presented elsewhere\cite{WSCMring}. Under some
conditions GM systems can also contain branch points \cite{branchmic}; we
forbid these in the present work.

After recalling some analytical predictions in Section~\ref{sec:Theory}
we discuss our new approach in detail in Section~\ref{sec:Simulation}.
With this algorithm we are able to vary the volume fraction over nearly three
orders of magnitude and we obtain equilibrated systems with average chain
length up to about $\Lav \approx 5000$ (see Tab.~\ref{tab:phiE}).
We present our computational results on static properties of EP without rings
in Section~\ref{sec:Results}.
A complete crossover scaling analysis ranging from the dilute regime
of swollen EP up to the dense Gaussian limit is performed;
for systems with large enough chains ($\Lav \gg 5$) we obtain
close agreement with recent analytical
predictions\cite{CatesCandau,Schaefer,Schoot}.
A delicate issue concerns finite-size effects which arise if one works at too
low a temperature for any given system size. These force the breakdown
of the (essentially) exponential MWD as one enters a state where a large
fraction of the monomers reside in a single chain.
Particular care is taken in this regard in Section~\ref{sec:FS}.
In the final Section~\ref{sec:Discussion} we summarize our findings.
Extension and development of these investigations to dynamic properties of
EP (without rings) will be reported in a companion paper \cite{WMCdynam}.

\section{Model and Analytical Predictions.}
\label{sec:Theory}
Before describing our computational investigation
we define the physical model, and recall some essential
analytical predictions concerning various properties of equilibrium
conformations of EP.

On a coarse-grained level (see Fig.~\ref{fig:sketchEP}) systems of EP are
characterized by the monomer volume fraction $\phi$,
the energy difference $E$ between saturated and unsaturated bond states,
the height of the activation energy barrier $B$,
the persistence length $l_p$ and the excluded volume size $b$
of the monomer (related to the cross sectional diameter of a giant micelle).
Additionally, parameters for the non-bonded interactions may be defined.

As approproate to the GM case we do not allow
\mbox{(i)}   closed rings or
\mbox{(ii)}  branching points and suppose that
\mbox{(iii)} all scission energies are independent
             of $\phi$ and position of the monomer within the chain.
Hence, the rates for scission and recombination $\ks=\exp(-(E+B)/k_BT)$ and
$\kr \propto \exp(-B/k_BT)$ are taken to be constant.
For GM simplification (iii) is expected to become accurate at high
salinity where electrostatic effects may be neglected\cite{CatesCandau}.

For EP that are long compared to the persistence length $l_p$,
the main departure from conventional theory of polymer
solutions\cite{deGennes79}
is that in micellar systems, the reversibility of the self-assembly process
ensures that the MWD $c(L)$ of the worm-like
polymeric species is in thermal equilibrium.
This contrasts with ordinary quenched  polymers for which the MWD
is fixed a priori, and equilibrium only applies to the remaining
(configurational) degrees of freedom. For EP, only the total volume fraction
of the system
\begin{equation}
\phi = \sum_{L}^{\Lmax} L c(L)
\label{eq:conserv}
\end{equation}
is a conserved quantity, rather than the entire distribution $c(L)$.
It is useful to introduce the normalized probability distribution
$\cN(L) = \left(\Lav/\phi\right) c(L)$, so that $\sum_{L} \cN(L) = 1$.
The maximum chain length \Lmax , given by the total number of monomers
in the system, becomes of relevance in the context of finite-size effects
(Sec.~\ref{sec:FS}).

At the level of a Flory-Huggins mean-field approximation (MFA) the grand
potential density $\Omega$ for a $d$-dimensional system of EP may be 
written as
\begin{equation}
\Omega =\sum_{L} c(L) \left[ \ln\left(c(L)b^d \right) + E + L \mu\right]
\label{eq:FMFA}
\end{equation}
where we choose energy units so that $k_BT=1$.
The first term is the entropy of mixing,
the second the scission energy $E$ of a bond and
the last term entails the usual Lagrange multiplier for the conserved
monomer density Eq.~(\ref{eq:conserv}).
Without loss of generality, we have suppressed in Eq.~(\ref{eq:F})
the part of the free energy linear in chain length.
Minimizing with respect to the MWD, paying attention to the constraint of
Eq.~(\ref{eq:conserv}), yields the exponential size distribution
$\cN(s)ds = \exp(-s) ds$
where the chemical potential defines a scaling variable $s=\mu L$.
This distribution has $\Lav \mu =1$, i.e. the scaling variable $s$
is given in the MFA by the reduced chain length $x\equiv L/\Lav = s$.
Using again
Eq.~(\ref{eq:conserv}) we then obtain the mean-chain length
\begin{equation}
\Lav = A \phi^{\grow} \exp(\delta E)
\label{eq:Lav}
\end{equation}
with a (nonuniversal) amplitude $A$ and the MFA-exponents
$\grow_{MF}=\delta_{MF}=1/2$. This result is expected to be a good
approximation near the
$\theta$-temperature and in the melt limit
where chains become free random walks uncorrelated with themselves and
their neighbors\cite{YA1}.

Note that the chain stiffness may in principle be incorporated in
the above description of flexible chains by adding to Eq.~(\ref{eq:FMFA})
a {\em free} energy term for the higher probability of trans-states. 
It is straightforward to work out, that for
reasonable bending energies this renormalizes the scission
energy $E$ by a constant of order unity (that is, of order $k_BT$). Here we
do not pursue stiffness effects and instead choose a persistence
length $l_p$ comparable to the monomer size $b$.

It is relatively simple to extend the above analysis to dilute and
semi-dilute solutions of EP.
We recall\cite{deGennes79} from standard polymer theory that
the correlation length \xiL\ for chains of length $L$
in the dilute limit is given by the size of the chain $\xiL = R \propto
L^{\nu}$.
When the chains (at given number density) become so long that they start to
overlap at $L \approx \Lstar \approx \phi^{-1/(\nu d -1)}$
the correlation length of the chain levels off and becomes the
(chain-length independent) size of a `blob' $\xiL = \xi \propto
\Lstar^{\nu}$. Here $d=3$ is the dimension of space, and
$\nu\approx 0.588$ is the swollen chain (self-avoiding walk) exponent.

The mean-field approach remains valid\cite{Cates88} so long as the basic
`monomer' is replaced by a blob of length \xiL. Across the entire
concentration range we may write the grand potential density as
\begin{equation}
\Omega = \sum_{L} c(L) \left[ \ln\left( c(L) \xiL^d \right)
                        - \ln \left((\xiL/b)^{d+\theta} \right)
                        + E + L\mu  \right].
\label{eq:F}
\end{equation}
We have taken into account here of the free energy change resulting from the
{\em gain in entropy} when a chain breaks so that the two new ends can
explore a
volume $\xiL^d$. This gain is enhanced by the fact that the excluded volume
repulsion on scales smaller than \xiL\ is reduced by breaking the chain;
this is accounted for by the additional exponent $\theta$.
Note that $\theta=(\gamma-1)/\nu \approx 0.3$ with $\gamma \approx 1.165$
(as we will confirm in Sec.~\ref{sec:MWD}).
In MFA $\gamma=1$,$\theta=0$, and Eq.~(\ref{eq:F}) simplifies to
Eq.~(\ref{eq:FMFA}), but this ignores correlations arising from excluded
volume effects. In the dilute regime Eq.~(\ref{eq:F}) can be rewritten as
\begin{equation}
\Omega =\sum_{L} c(L) \left[ \ln(c(L)b^d) + (\gamma-1)
\ln(L) + E + L
\mu\right]
\end{equation}
so that the relation to the well-known partition function of
self-avoiding walks (with an effective coordination number $\tilde{z}$),
$Q_L \propto \tilde{z}^L L^{\gamma-1}$, is recovered.

Note that in concentrated and semidilute systems there are minor corrections
to Eq.~(\ref{eq:FMFA}) from the small fraction of chains that
are too long for their excluded volume interactions to be screened by
the surrounding chains (under melt conditions, this
applies\cite{deGennes79} for those chains whose length exceeds $\Lav^2$).
This contribution is exponentially small and will be neglected. Note
however that the contribution of short chains (smaller than or comparable
to the blob size) in the semidilute regime is properly included in
Eq.~(\ref{eq:F}), so long as the $L$-dependence of $\xiL$ is taken into
account. In general, this dependence is contained in a scaling function
\begin{equation}
\xiL(s)  = \xiL(s/s^*)
\label{eq:Xs}
\end{equation}
Where the scaling variable $s=L\mu$ was already introduced above,
and $\sstar=\Lstar\mu$. This function asymptotes to the radius of gyration
and the blob size at large and small $s$ respectively.

Minimization of Eq.~(\ref{eq:F}) at fixed $\phi$ yields the MWD
\begin{equation}
\cN(L) dL = \cN(s) ds \propto \xiL(s) \exp(-s) ds
\label{eq:ps}
\end{equation}
which depends via \xiL\ on $s^*$.
We see that the effective exponent
$\gammaeff(\sstar) \equiv s/x = \Lav \mu= \left< s \right> =\int s p(s) ds$
is not in general equal to unity.
Hence, the scaling variable $s$ is not
necessarily given by the reduced chain length $x$ as in MFA and
as supposed in ref.~\cite{Guj}, but by $s=\gammaeff x$.
In the two limits far away from the crossover line one can readily
calculate the integrals. In the semi-dilute limit ($\sstar \ll 1$)
we obtain $\Lav \mu=\gammaeff = 1$, i.e. as in the MFA $s=x$,
but in the dilute limit ($\sstar \gg 1$) we get
$\Lav \mu = \gammaeff = \gamma$, i.e. $s=\gamma x$.
Substituting everything we obtain finally the distributions
\begin{equation}
\cN(x) dx =
\left\{  \begin{array}{ll}
\exp(-x) dx & \mbox{($\Lav \gg \Lstar$)} \\
\frac{\gamma^{\gamma}}{\Gamma(\gamma)} x^{\gamma-1} \exp(-\gamma x) dx
        &\mbox{($\Lav \ll \Lstar$)}
\end{array}
\right.
\label{eq:p}
\end{equation}
We remark that any observed breaking of the scaling
$\cN(L)dL = \cN(x)dx$ is an indication of either crossover between
both density regimes ($\Lav \approx \Lstar$) or
finite-size effects ($\Lav \approx 1$ or $\Lav \approx \Lmax$).
One expects for semi-dilute configurations close to the
crossover line some reminiscence of the dilute behavior,
which should show up in a slightly higher \gammaeff.

As in the MFA-case we obtain a mean chain length 
\begin{equation}
\Lav =  \Lstar (\phi/\phistar)^{\grow}
\label{eq:Lscal}
\end{equation}
which may be cast in the generic form Eq.~(\ref{eq:Lav}), but with
the exponents $\grow_d=\delta_d=1/(1+\gamma)\approx 0.46$ in the dilute
and $\grow_s=1/2 (1+(\gamma-1)/(\nu d-1) \approx 0.6$, $\delta_d =1/2$ in the
semi-dilute limit. 
Thus the concentration dependences of the mean molecular weight \Lav\
in the dilute and semidilute limits differ slightly form the one predicted
by simple mean-field theory. 
Here finally we {\em define} ($L^*,\phi^*$) quantitatively as the coordinates
of the intercept of the dilute and semidilute asymptotes on a plot of \Lav\
vs. $\phi$ (see Fig. \ref{fig:L} below).
Accordingly we may write \begin{eqnarray}
\phistar & = & P \exp(-E/\varphi) \nonumber\\
\Lstar & = & Q \exp(E/\kappa)
\label{eq:phiLstar}
\end{eqnarray}
with exponents
$\varphi = (\grow_s-\grow_d)/(\delta_s-\delta_d)
         = 1 + (\gamma+1)/(\nu d -1) \approx 3.8$ and
$\kappa  = (\nu d -1 ) \varphi \approx 2.93$.
The amplitudes $P$ and $Q$ are similary related to the prefactors
$A_d$ and $A_s$ of the mean chain length (as defined in Eq.~(\ref{eq:Lav})
in each density regime  (see also Eq.~(\ref{eq:PQ}) below).

We consider finally the specific heat capacitly of the system. This offers a
possible experimental measure of the typical scission energy
$E$. Assuming this to be purely enthalpic and independent of temperature, the
internal energy density $U$ given by the density of end monomers
$U = E \sum_{l} c(l) = E \phi/ \Lav$. From this we get the specific heat
per monomer
\begin{equation}
\CV = \frac{1}{\phi} \frac{\partial U}{\partial T} = \delta \frac{E^2}{\Lav}
= \frac{\delta}{A \phi^{\grow}} E^2 \exp(-\delta E).
\label{eq:CV}
\end{equation}
One verifies that $\CV$ has a maximum at $E = 2/\delta$ which shifts slightly
to higher values at lower densities (from 4 above \Lstar\ to 4.4 below).

\section{The Algorithm and Configurations.}
\label{sec:Simulation}

\begin{figure}
\centerline{\epsfysize=6cm
\epsfig{file=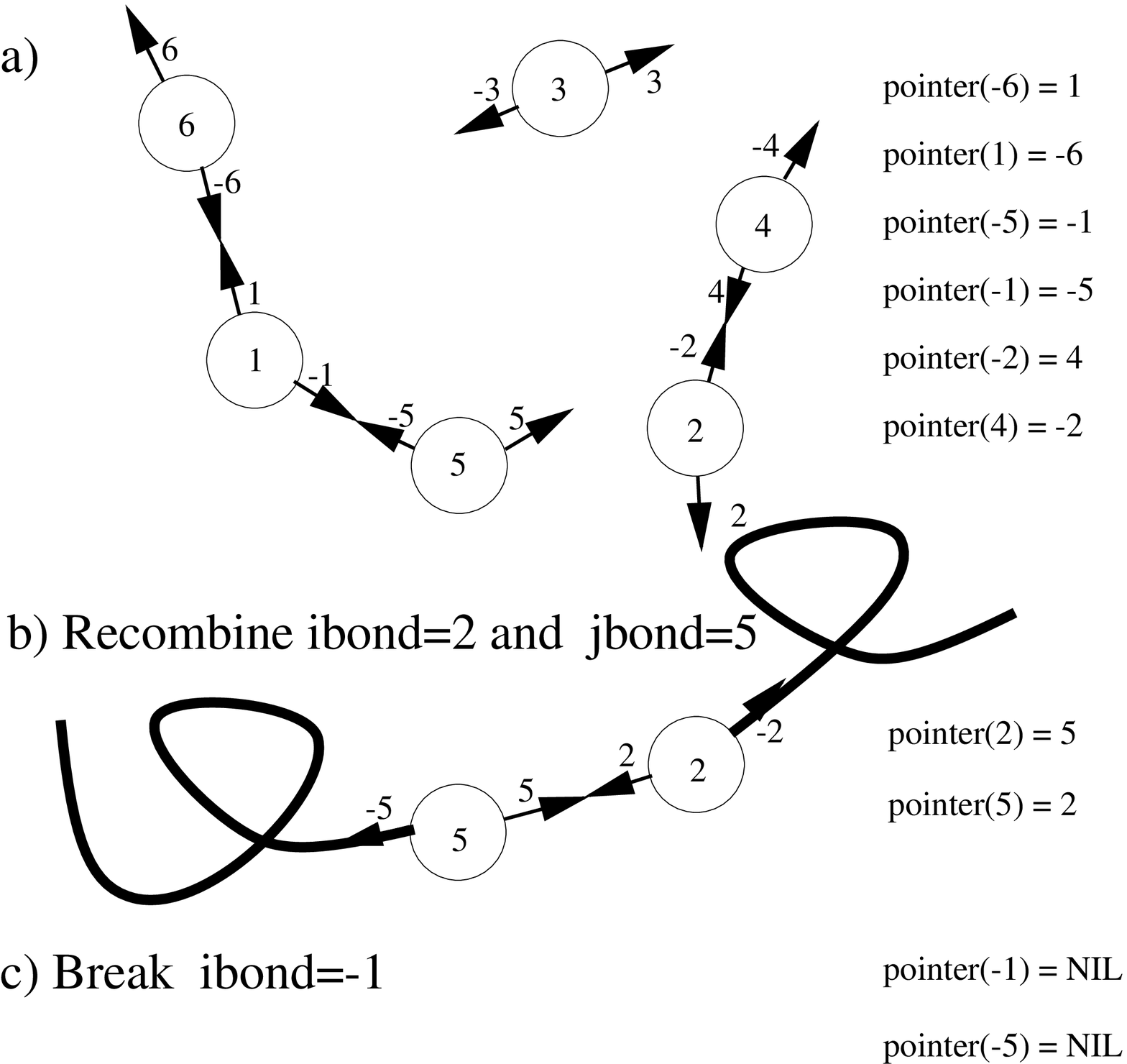,width=60mm,height=40mm,angle=0}}
\vspace*{0.4cm}
{\small FIG.~2. 
Sketch of algorithm:
(a) Each monomer has two (saturated or unsaturated) bonds.
Chains consists of symmetrically connected lists of bonds:
{\em jbond = pointer(ibond)} $\Leftrightarrow$ {\em ipoint = pointer(jpoint)}.
The pointers of all end-bonds point to \mbox{NIL}.
(b) Recombination of two initially unsaturated bonds {\em ibond = 2}
and {\em jbond = 5} connects the respective monomers {\em imon=2} and {\em
jmon=5}.
Note that only two pointers have to be changed and that the remaining
chains (bold lines) behind both monomers are not involved.
(c) Breaking a saturated bond {\em ibond} requires resetting the pointers of
the two connected bonds {\em ibond} and {\em jbond = pointer(ibond)} to
\mbox{NIL}.
\label{fig:sketchAlgo}}
\end{figure}

In EP systems bonds between monomers break and recombine constantly
and the chains are only transient objects (Fig.~\ref{fig:sketchEP}).
Therefore it is relatively inefficient to base the data structure on the
{\em chains} (such an approach penalized either by sorting times or waste of
memory).  Rather, one has to base it on the {\em monomers},
or even better on the two saturated or unsaturated {\em bonds} of each monomer.
This brings the algorithm as close as possible to what actually
happens in systems of EP and makes it possible via pointers between bonds
to avoid all sorting procedures, time consuming nested loops and arrays,
at the expense of only one additional list required.
Our chosen data structure is explained for three initial chains in
Fig.~\ref{fig:sketchAlgo}, one
of which is actually a free monomer with two unsaturated bonds.

Using the assumption that no branching of chains is allowed, the two
(possible) bonds of each monomer $imon$ are called $ibond=imon$ and
$ibond=-imon$. No specific meaning (or direction) is attached to the sign:
this is merely a convenience for finding the monomer from the bond list
($imon=|ibond|$).
Pointers are taken to couple independently of sign,
see Fig.~\ref{fig:sketchAlgo}.
In the proposed algorithm the bonds are coupled by means of a pointer list
in a completely transitive fashion
($jbond=pointer(ibond) \Leftrightarrow ipoint=pointer(jpoint)$)
to make recombinations and scissions as fast as possible;
only two simple vector operations are required for breaking bonds or
recombination as shown in  Fig.~\ref{fig:sketchAlgo}.
Note that this would be impossible in any algorithm involving only one bond
per monomer; the latter requires an implicit sequential order of segments in
the chains, forcing sorting operations of the order of the mean-chain length
for every recombination. Unsaturated bonds at chain ends
point to \mbox{NIL}. Only these bonds may recombine. With this algorithm, no
explicit distinction between end-monomers, free monomers or middle monomers
is required.

As a work-horse, harnessed to this new data organization, we exploit the {\em
bond-fluctuation model} (BFM)\cite{BFM} for the DMC simulation of polymers.
This choice was mainly motivated by its efficiency, and the large amount
of existing data available on monodisperse conventional polymers which can
serve as a reference against which to compare the present work.
However, we emphasize that a different choice,
such as an off-lattice MC algorithm\cite{AndreyOfflattice}, could equally
well be combined with our data structure.

In the BFM an ``effective monomer'' consists of an elementary cube whose
eight sites on the hypothetical cubic lattice are blocked for further
occupation\cite{BFM}. We consider the formulation of the BFM on a dual
lattice, i.e. the center of an effective monomer is represented by {\em one}
site on a simple cubic lattice as introduced by M.~M\"uller \cite{Marcus}.
Excluding all $26$ neighboring sites from occupancy renders the model
equivalent to the original model proposed by Carmesin and Kremer\cite{BFM}.
The volume fraction $\phi$ is the fraction of lattice sites
blocked by the monomers.
The monomers of a polymer chain are connected via bond vectors $\bf b$,
which are taken from  the allowed set
$P(2,0,0), P(2,1,0), P(2,1,1), P(3,0,0)$, and $P(3,1,0)$,
where $P$ stands for all permutations and sign combinations of coordinates.
A bond corresponds physically to the end-to-end distance of a group of $3-5$
successive monomers and can therefore {\em fluctuate} within some range of
lengths. All length are measured in units of the lattice spacing $a$.
The algorithm combines typical advantages of the lattice MC methods with
those from the continuous Brownian dynamics algorithm. As defined above, it
corresponds to good solvent conditions without hydrodynamics (but
respecting entanglement constraints)\cite{BFM,Wolferl}.

As explained above (Sec.~\ref{sec:Theory}) the ends of a given EP
are not allowed to bind together in the presented study;
before every recombination we have therefore to check that both monomers do
not belong to the same chain.
Because there is no direct chain information in the data structure
this has to be done by working up the list of links (which adds only four
lines to the source code). In physical time units the simulation becomes {\em
faster} for higher $E$:
the number of recombinations per unit time goes down like $\exp(-E)$,
but the chain length only up as $\exp(E/2)$.

The barrier energy $B > 0$ is taken into account by setting an
attempt frequency \freqB\ for scissions and recombinations. This is a
convenient tool for testing dynamic behavior of the system at different
lifetimes of the chains\cite{WMCdynam}, although for static properties as
studied in this paper, the choice of $B$ is immaterial. Therefore in almost
all runs reported here we set $B=0$.  Those sites of the lattice that are not
occupied by monomers are considered empty (vacancies) and contribute to the
free volume of the system. We may in principle assign an energy $-w$ ($w > 0$) 
for the nonbonded interaction between monomers in the system\cite{YA1},
and a bending energy $s \sin(\theta_{ij})$ with
$\theta_{ij}$ being the angle between consecutive bonds\cite{WPB1}.
In the present investigation, however, we focus exclusively on the
process of equilibrium polymerization of entirely flexible chains in
an athermal solvent setting
$w=s=0$.

Time is measured, as usual, in Monte Carlo steps (MCS) {\em per monomer}.
Each MCS is organized as follows:
\begin{itemize}
\item{ A monomer is chosen at random and allowed to perform
a move according to the BFM algorithm\cite{BFM}.}

\item{With a frequency $\freqB \equiv \exp(-B)$, i.e. every $1/\freqB$ MCS,
one of the bonds is chosen at random (remember that there are twice
as many bonds as monomers).
If one of the bonds happens to be a saturated $P(2,0,0)$-bond
an attempt is made to break it, otherwise if it is unsaturated,
i.e. the monomer is at the end of a chain or a free monomer,
an attempt is made to create a bond with another monomer that
might be present on {\em any} of the 6 neighboring $P(2,0,0)$ sites.}
Applying the Metropolis algorithm\cite{Binderbook79}
a scission is performed whenever the value of a random number between 0 and 1
is smaller than $\exp(-E)$. On the other hand, with the bond energy being
positive ($E>0$), recombination is always accepted so long as a ring is
not thereby created.
\end{itemize}

Note that $P(2,0,0)$-bonds are broken irrespective of which particular
bond vector, i.e. which of the $6$ possibilities, they stand for, and that,
therefore, detailed balance requires that for recombination {\em all}
(and not just {\em one} as in ref.~\cite{YA1}) of these sites
have to be checked for possible unsaturated bonds.
These neighboring sites have to be checked {\em randomly};
a typewriterlike search along the list of possible bond vectors
creates correlations in violation of the detailed balance requirement.
(This has subtle but measurable consequences.) Note finally that our
decision to restrict the breaking and recombination to the
6 shortest ($P(2,0,0)$) bonds avoids ergodicity problems arising from
crossed pairs of bonds that can result in immobile monomers.\cite{shortbonds}

\begin{figure}
\centerline{\epsfysize=6cm
\epsfig{file=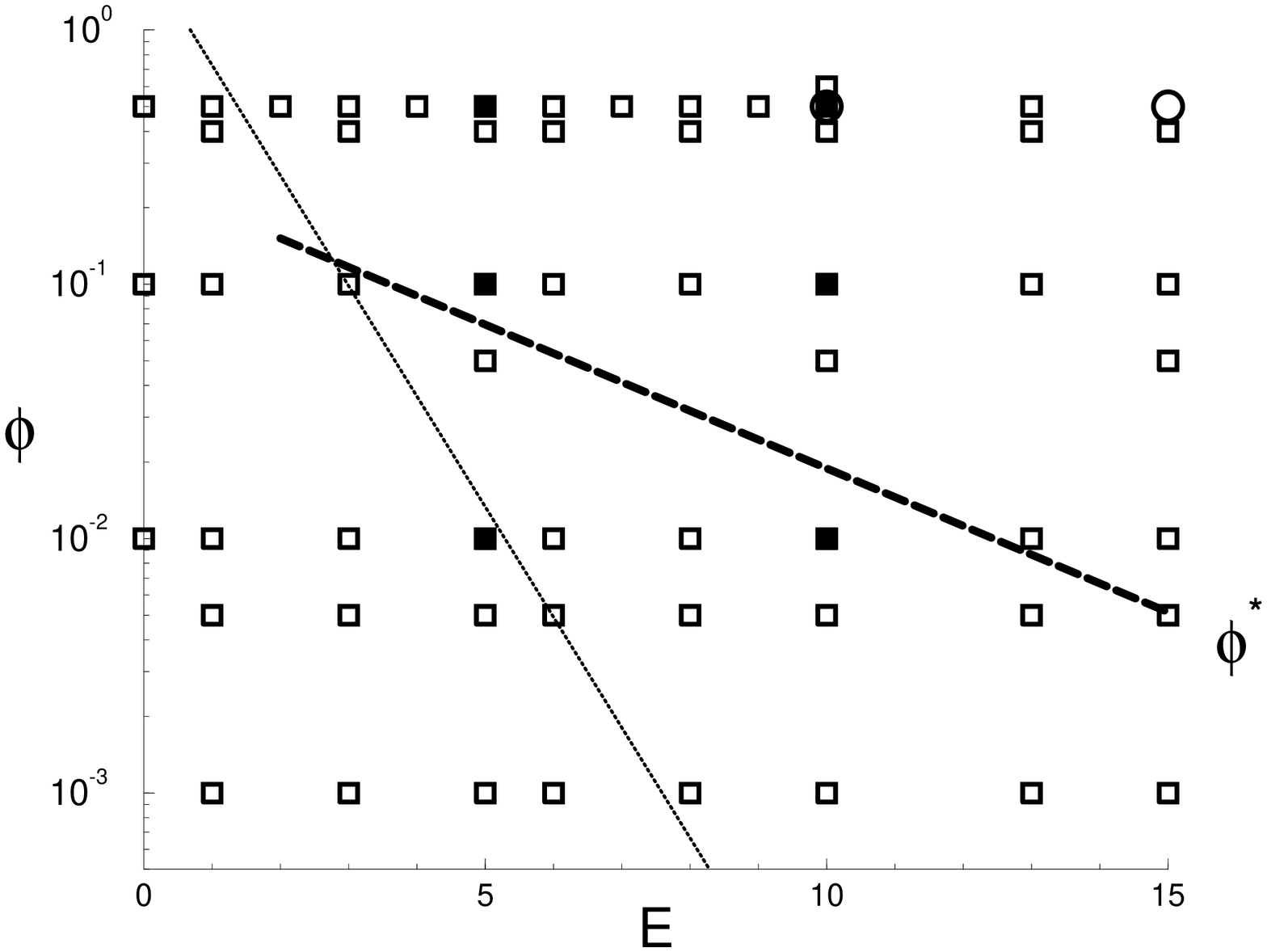,width=70mm,height=90mm,angle=-90}}
\vspace*{0.4cm}
{\small FIG.~3. 
Simulation parameters $(E,\phi)$ in relation to the crossover
density
$\phistar$. Above and to the right of the dashed line is the semi-dilute
regime, below and to the left the dilute regime.
The crossover density follows an exponential decay with
$\phistar \approx P \exp(-E/\varphi)$ where $\varphi=3.82$ and
$P=0.26$ consistent with Eqs.~(\ref{eq:phiLstar}) and (\ref{eq:PQ}).
At volume fractions of about $\phi=0.5$ the correlation length
becomes of order of the monomer size (the melt regime).
The dotted line corresponds to $\Lav=5$; to the left of this line
chains are too short for good scaling.
For some systems (filled symbols) we have in addition systematically varied
the frequency of scission/recombination $\freqB$ by varying the activation
energy Ref.~\protect\cite{WMCdynam}.
For certain parameter choices (circles) we have checked carefully for
finite-size effects.
\label{fig:phiE}}
\end{figure}

In the presented study the volume fraction was varied over nearly three orders
of magnitude from $\phi=0.001$ ($\Lmax=1,000$ monomers per box) to $\phi=0.6$
($\Lmax=75,000$ monomers per box) as shown in Fig.~\ref{fig:phiE}.
For densities smaller than $\phi=0.1$ cubic lattices with linear dimension
$\Lbox = 200$ were used, for higher densities a smaller box with
$\Lbox = 100$. This should be compared to ref.~\cite{YA1} where a
$\Lbox=30$-box with $\Lmax=1,300$ particles ($\phi = 0.4$) was the largest
achievable. One should bear in mind that densities around $0.5$ correspond to
extremely dense systems (melt conditions) in the BFM, since at higher
densities the blocking of neighboring sites by other
monomers leaves no room for movement and the system goes effectively into
a glassy state\cite{BFM,Wolferl}.

The starting configuration consists of randomly distributed and nonbonded
monomers which we cool down step by step (a sequence of so-called `T-Jumps')
each step sampling a higher scission energy up to the maximum $E = 15$.
As mentioned above, for the static results we usually set $B = 0$, i.e.
$\freqB = 1$. Due to the constant breaking and recombining of the bonds
the equilibration is then much faster than in systems of quenched polymers.
Following Rouault et al\cite{YA2} we have checked this explicitly by
monitoring the evolution of mean-chain length $\left< L(t)\right>$ and
radius of gyration $\left< R_g^2(t) \right>$ after every T-Jump.

Measurements of static properties such as the mean-square end-to-end
distance \Ree, the radius of gyration \Rgg\ or the specific heat \CV\
(see Tab.~\ref{tab:phiE}) were performed in intervals of roughly
$\tau_t/10$,
where $\tau_t$ is the terminal relaxation time of the system for the parameters
$(\phi,E)$ considered\cite{WMCdynam}. Typical runs (with  $B = 0$) covered
easily up to $10-100 \ \tau_t$.
The melt density $\phi=0.5$ was a particular focus for our study of
dynamical properties\cite{WMCdynam}, requiring better statistics.
Hence, for this $\phi$ we have sampled over 128 independent
configurations. However, for all static properties other than the specific
heat \CV\ and, for the largest $E$ values some MWD data, a long run from
a single starting configuration was quite sufficient. Indeed, statistical
errors are generally within the symbol size in the data presented below.

There are two different kinds of ``finite-size effects" in this simulation.
The first, and most important, arises from chains that are too small.
Not surprisingly, we found in configurations with $\Lav < 5$
(below the dotted line in fig.~\ref{fig:phiE}) non-universal behavior.
For clarity these data points are omitted in most of our plots.
Second, for systems with (essentially) exponential MWD
one has to worry about the system size whenever \Lav\ becomes of the
order of the total number of monomers in the box \Lmax.
This important issue will be addressed in Section~\ref{sec:FS}
where we present results of a systematic finite-size study
summarized in Tab.~\ref{tab:FS} confirming unambiguously
that the systems reported are indeed sufficiently large.


Most of the simulations on static properties were performed on two single
DEC Alpha workstation over a period of year.
A parallel version of the algorithm, however, was also developed
and some of the computations at melt density mentioned above
have been carried out on the facilities of EPCC, Edinburgh.
The latter computations focused mainly on the dynamics
of equilibrium polymers as we report elsewhere \cite{WMCdynam}.

\section{Computational confirmation of scaling predictions.}
\label{sec:Results}

In what follows we examine the influence of density $\phi$ and
scission energy $E$ on mean chain length \Lav, MWD $c(L)$,
specific heat per monomer \CV\ and
the size of the chains, as measured e.g. by the radius of gyration \Rgg.

\subsection{Mean Chain Length.}
\label{sec:L}

\begin{figure}
\centerline{\epsfysize=6cm
\epsfig{file=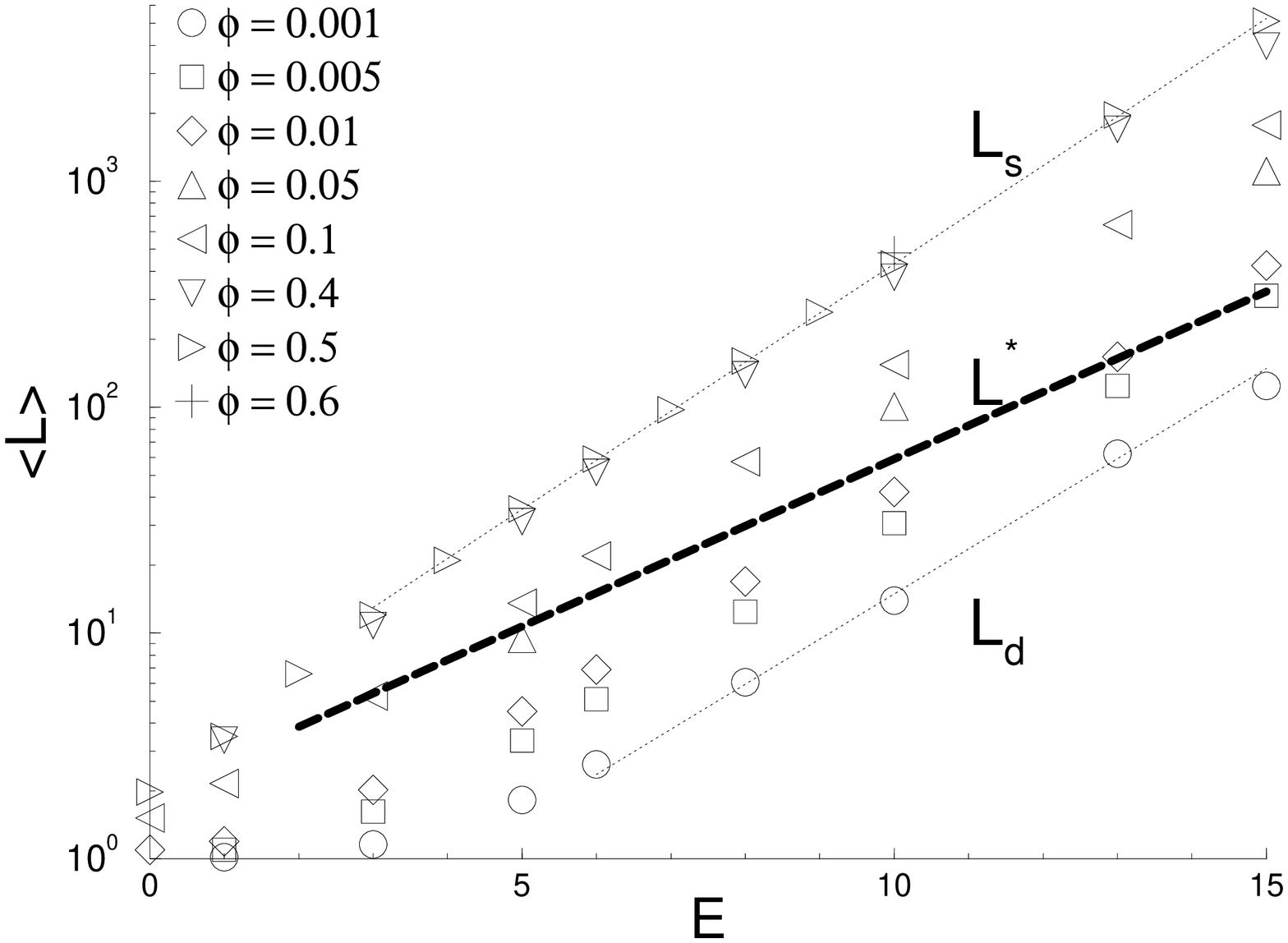,width=50mm,height=60mm,angle=-90}}
\centerline{\epsfysize=6cm
\epsfig{file=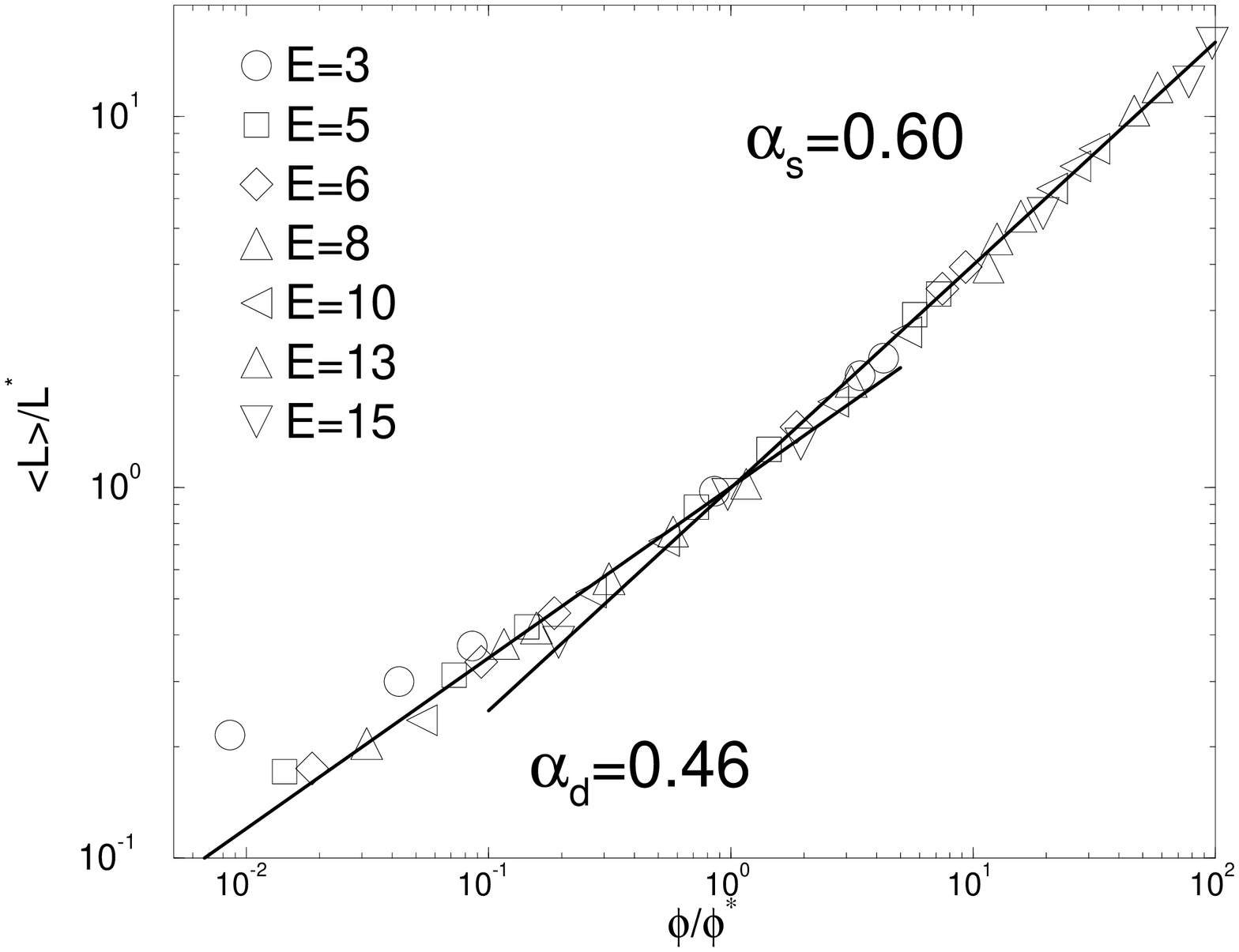,width=50mm,height=70mm,angle=-90}}
\vspace*{0.4cm}
{\small FIG.~4. 
(a) The average chain length $\Lav$ for a wide range of
densities $\phi$ and energies $E$. (b) Rescaled average chain length
$\NB=\Lav/\Lstar$ versus reduced density $\phi/\phistar$ confirming
the scaling Eq.~(\ref{eq:Lscal}) and the amplitudes Eq.~(\ref{eq:PQ}).
The two slopes are comparisons with the predicted growth exponents
$\grow_d \approx 0.46$ and $\grow_s \approx 0.6$.
\label{fig:L}}
\end{figure}

In Fig.~\ref{fig:L}a we show in a semi-log plot the measured variation
of the mean chain length $\Lav$ versus the bond energy $E$.
Note that we have been able with this new algorithm to obtain mean chain
length of up to $\Lav \approx 5,100$  (see Tab.~\ref{tab:phiE})
which is comparable with (at least some) experimental systems of EP.
Giant micelles, for example, are somewhat rigid and the persistence length
relatively large,
$l_p \approx 16\; nm$ whereas $\Rgyr \approx 100\; nm$,
resulting in around 100 statistical segments\cite{Berret}, although
much longer chains could arise in some semidilute systems \cite{CatesGranek}.
As mentioned in the Sec.~\ref{sec:Simulation} and elaborated further
in Sec.~\ref{sec:FS} below,
it is the finite size of the systems, rather than the equilibration time,
that prevents us studying higher $E$ (lower temperatures)
since the largest chain would then comprise too high a proportion of
the total available monomers in the system.

For long enough chains (above the dotted line in Fig.~\ref{fig:phiE})
the chain length increases exponentially with scission
energy $E$, as predicted by Eq.~(\ref{eq:Lav}). The data confirm with high
precision the predicted exponents
$\delta_d \approx 0.46 \pm 0.01$ and $\delta_s=0.5 \pm 0.005$
in the dilute and semidilute regimes respectively.
The growth exponents $\grow$ are most readily
confirmed (following the scaling prediction Eq.~(\ref{eq:Lscal}))
by directly plotting the ``number of blobs" $\NB=\Lav/\Lstar$
against the reduced density
$\phi/\phistar$. The data collapse onto a single master curve, seen in
Fig.~\ref{fig:L}b, is indeed remarkable and is one of the main results of
this work. The two indicated slopes are comparisons with the two
asymptotically predicted growth exponents ($\grow_d = 0.46$, $\grow_s \approx
0.6$).  (Note that in the dilute limit finite-size effects are visible for
low $E$ values where chains are extremely short ($\Lav < 5$).)
This finding is at variance with ref.\cite{Kro} where a much stronger
growth with density is reported for systems containing only $8400$ monomers
with mean-chain lengths up to $210$.

By plotting $A=\Lav/\phi^{\grow}/\exp(\delta E)$ versus \Lav\ or $E$ one
fits for each density regime an amplitude $A_d \approx 3.6 \pm 0.2$
and
$A_s \approx 4.4 \pm 0.1$ respectively.
From these prefactors the amplitudes governing \phistar\ and \Lstar\ defined
in Eq.~(\ref{eq:phiLstar}) can be obtained
\begin{eqnarray}
P & = & \left( A_d/A_s \right)^{1/(\grow_s-\grow_d)} \approx 0.3 \pm 0.1
\nonumber \\
Q & = & A_s P^{\grow_s} \approx 1.9 \pm 0.5
\label{eq:PQ}
\end{eqnarray}
(The high error bars are caused by the small difference in the
two growth exponents resulting in the large exponent $1/(\grow_s-\grow_d)$.)
Although these are defined by the crossing point of the two asymptotes in
Fig.~\ref{fig:L}b, it is notable that the asymptotes are followed, within
numerical accuracy, all the way to the crossing point. (A large deviation
is of course not possible since the two asymptotic slopes are not very
different.) Thus the crossover between the two regimes occurs rather sharply
at  $\Lav/\Lstar=\phi/\phistar=1$, and this can be used to define a
crossover line on the $\phi,E$ or $\Lav,E$ plot.  Indeed dashed lines in
fig.~(\ref{fig:phiE}) and fig.~(\ref{fig:L}a) indicate the position of the
crossover lines \phistar\ and \Lstar\ using the exponents $\varphi=3.84$ and
$\kappa=2.93$ together with the above amplitudes $P$ and $Q$.

\subsection{Molecular Weight Distribution.}
\label{sec:MWD}

We consider first the MWD in the dilute ($\NB \ll 1$) and
semidilute ($\NB \gg 1$) limits, far away from the crossover.
Thereafter we discuss the crossover effects near
$\NB \approx 1$ where a significant fraction of the chains are still
smaller than the  blob size of the semidilute network.

\begin{figure}
\centerline{\epsfysize=6cm
\epsfig{file=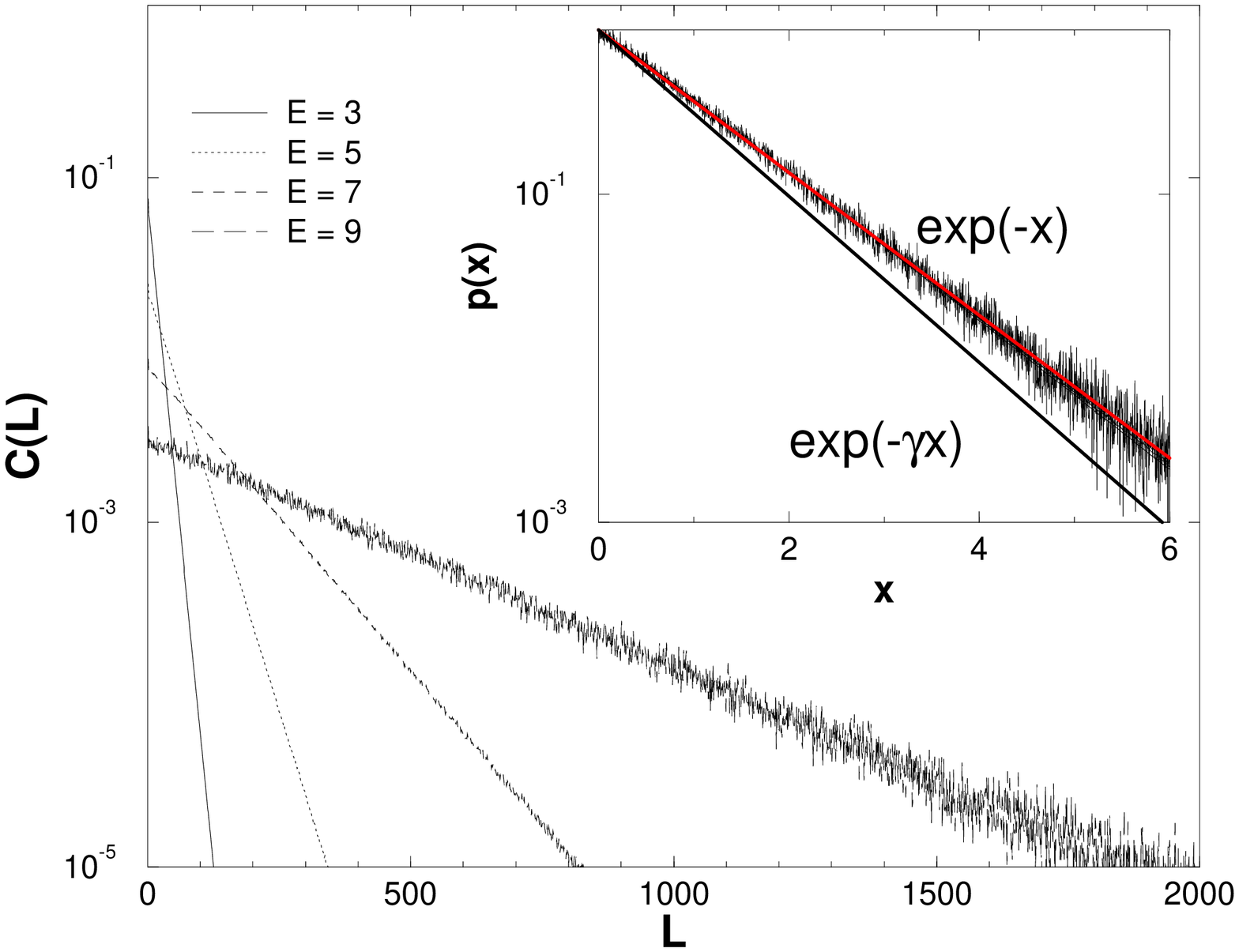,width=55mm,height=80mm,angle=-90}}
\centerline{\epsfysize=6cm
\epsfig{file=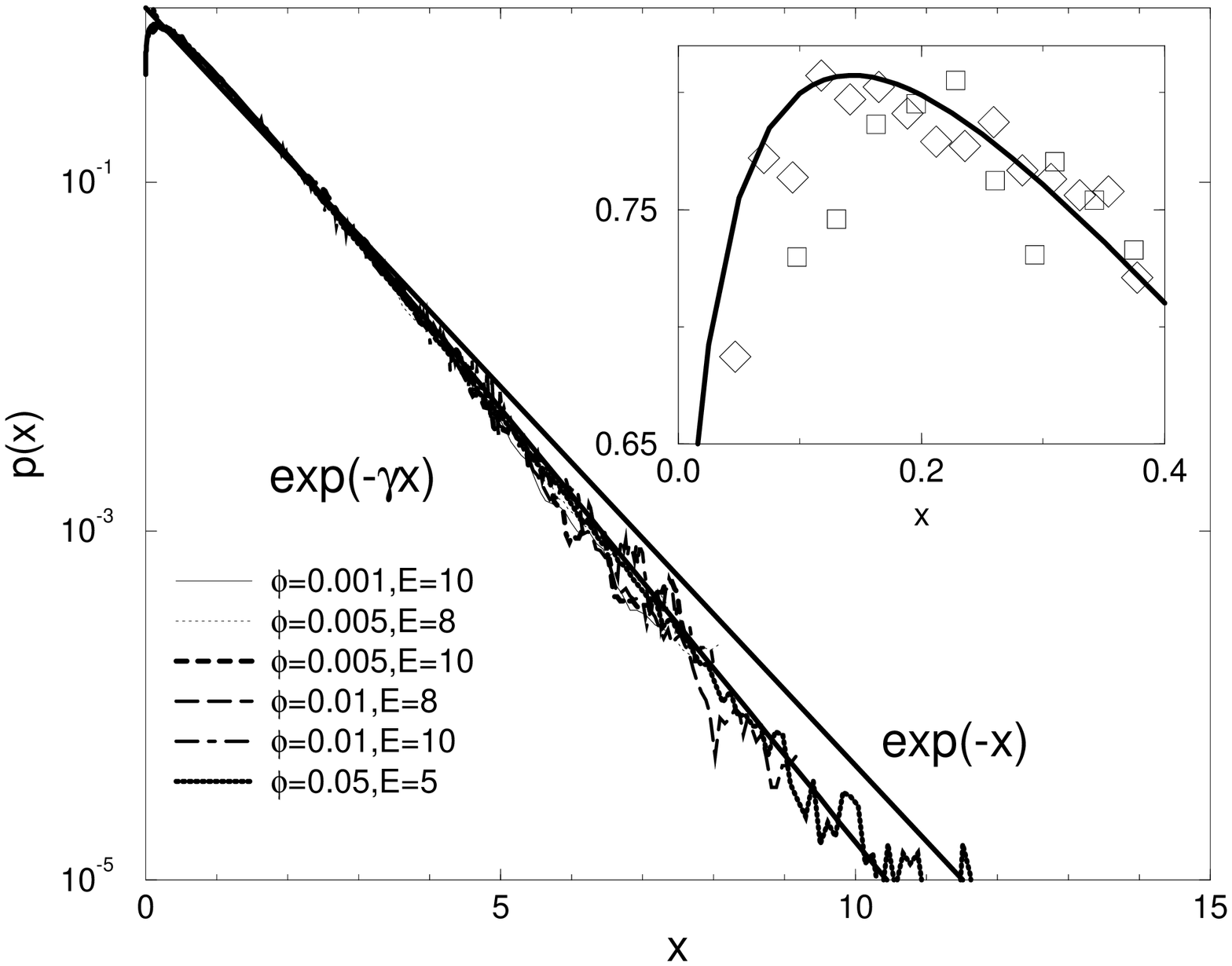,width=55mm,height=80mm,angle=-90}}
\vspace*{0.4cm}
{\small FIG.~5. 
Molecular Weight Distribution.
(a) $c(L)$ at high (melt) density $\phi=0.5$ for energies $E$
as indicated in the figure.
Insert: Data collapse of the normalized distribution
$\cN(L)dL=\cN(x) dx = \exp(-x) dx$.
(b) Dilute limit confirming $\cN(x) \propto \exp(-\gamma x)$.
Insert: MWD for small $x$ for two configurations
with $\phi=0.005$ (squares) and $\phi=0.01$ (diamonds) both at $E=10$ in
the dilute regime.
We compare (bold line) with the prediction of Eq.~(\ref{eq:p}b).
\label{fig:plowhigh}}
\end{figure}

The distribution of chain lengths at equilibrium, $c(L)$, is
presented in Fig.~\ref{fig:plowhigh}a on semi-log axes
for various $E$ at high density $\phi=0.5$.
The fluctuations in the sampled lengths increase considerably
for very long chains where correlations between successive
configurations deteriorate the statistics.
To the available accuracy, the normalized distribution $\cN(x)$ plotted
versus the reduced chain length
$x=L/\Lav$ collapse perfectly on single `master' curve as
shown in the insert of Fig.~\ref{fig:plowhigh}a; thus
the mean chain-length $\Lav$ contains all energy information.
The exponential decay confirms Eq.~(\ref{eq:p}a).
For comparison we have indicated the $\exp(-\gamma x)$ behavior which
is clearly not compatible with the data.
This finding is in agreement with ref.~\cite{YA3},
but in clear contrast to Gujrati~\cite{Guj} according to whom
the Schulz distribution holds independently of the overlap.

While at high densities we observe perfect exponential scaling of $\cN(x)$,
at lower dilute densities (with sufficiently long chains) our results are
qualitatively consistent with the Schulz distribution Eq.~(\ref{fig:plowhigh}b).
We compare in Fig.~\ref{fig:plowhigh}b this prediction (bold line) with
data sets for configurations in the dilute limit.
To stress the systematic difference, we have included the high density
prediction.

As shown in the insert of Fig.~\ref{fig:plowhigh}b our MWD at dilute
densities
are also qualitatively
consistent with the additional power-law dependence $\cN  \propto x^{\gamma-1}$
in the limit of small $x$. Note that the maximum of the distribution
is at $x_M = (\gamma-1)/\gamma \approx 0.1$ corresponding to a chain length
$L = 0.1 \Lav \approx 3.1$ for $\phi=0.005$ and $\approx 4.2$ for $\phi=0.01$.
Hence, we could not expect to reproduce accurately the power-law regime
for $x \ll x_M$. This would require configurations of at least $\Lav
\approx 1000$
in the dilute regime; due to the finite-size effects in the range of
scission energies used (discussed in Sec.~\ref{sec:FS} below) this is at
present not feasible.
(Since in $2d$ one has $\gamma = 43/32$ which is much farer from the MFA
value
$\gamma=1$ this part of the distribution can be more efficiently probed in
$2d$, as done recently  by Rouault and Milchev\cite{YA3}.)
Qualitatively, however, we believe that this result is unambiguous.
Note that we see no evidence for a possible {\em negative} exponent in the
power law in Eq.(\ref{eq:p}),
as postulated in some treatments of the unusual diffusive behavior in GM
\cite{Bouchaud}.

\begin{figure}
\centerline{\epsfysize=6cm
\epsfig{file=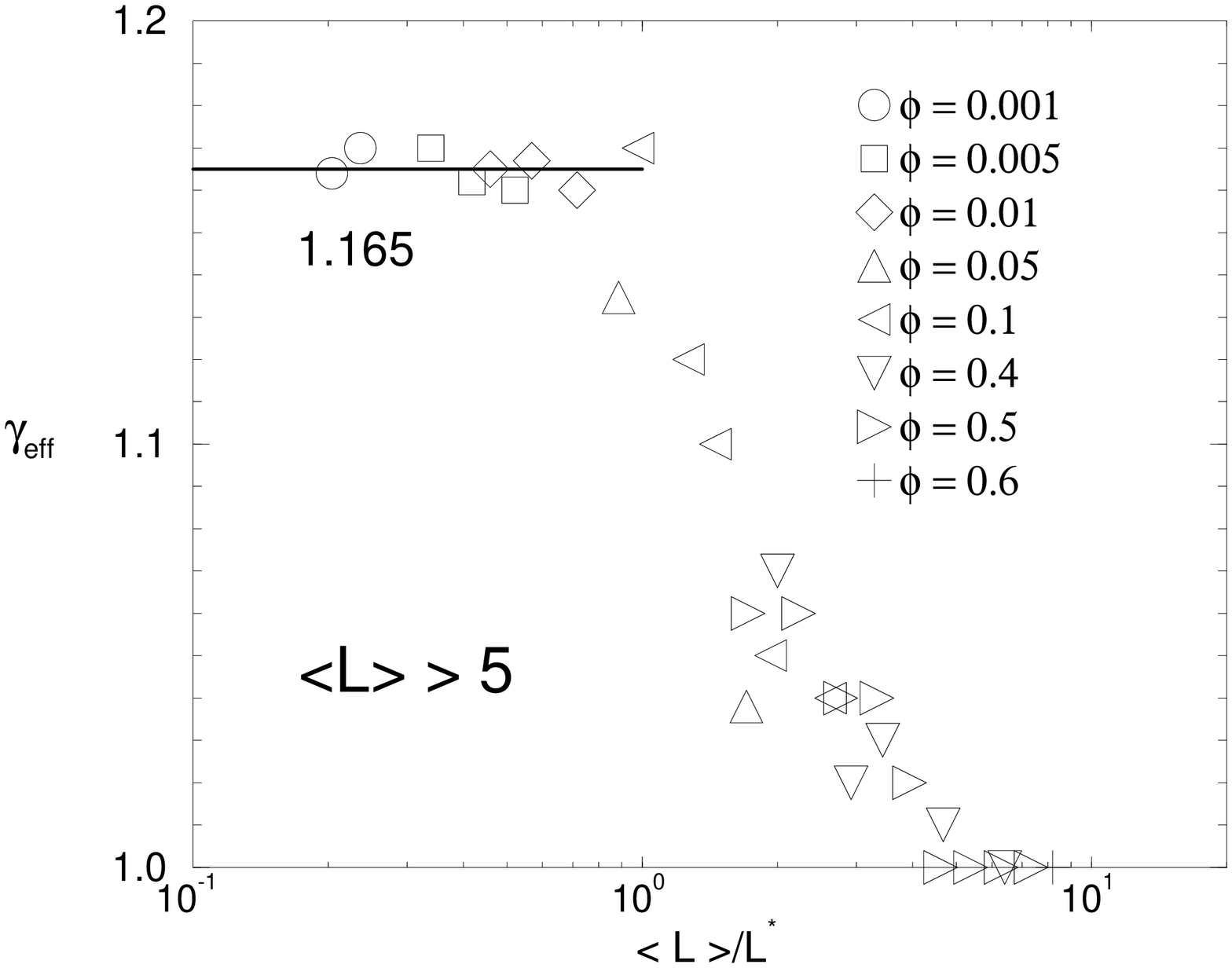,width=55mm,height=80mm,angle=-90}}
\vspace*{0.4cm}
{\small FIG.~6. 
Effective exponent $\gamma_e$ versus the number of blobs per chain
$\NB=\Lav/\Lstar$.
\label{fig:gammaeff}}
\end{figure}

At intermediate densities, slightly above the crossover
line, a non-negligible fraction of chains are smaller than the blob size, and
are thus fully swollen (these chains may fit among the network of
chains of average size $\Lav$ without being seriously perturbed by
the interchain interaction).  The distribution therefore crosses over
smoothly from the dilute limit  as depicted in Fig.~\ref{fig:plowhigh}b
to the semi-dilute limit of  Fig.~\ref{fig:plowhigh}b (data not shown).
Throughout the parameter range we have fitted systematically the effective
exponent \gammaeff\ from the $\exp(-\gammaeff x)$-tail of the MWD.
The values obtained are plotted in Fig.~\ref{fig:gammaeff} versus the
number of blobs \NB.
We confirm $\gamma_e \rightarrow \gamma=1.165$ and $\gamma_e \rightarrow
\gamma=1$
in the dilute and semi-dilute limit respectively. In between we observe a
crossover for the effective exponent. Note that the error bars are mostly
around $\pm 0.01$, however, a much higher accuracy is in principle
feasible by this method. Note that the value obtained in the dilute limit
compares well with the best renormalization group estimate
$\gamma = 1.1615\pm 0.0011$\cite{desCloizJannink}. We believe this is the
most reliable simulation determination so far of this well-known polymer
exponent.

In passing, we recall that the polydispersity index
$I=\left<L^2\right>/\Lav^2$, i.e. the ratio of weight average and
number average molecular weights, becomes $I=1+1/\gamma$ and $I=2$
in the dilute and semidilute limits respectively. This offers (in principle)
an additional method to check the $\gamma$-exponent.
However, due to the difficulty to measure accurately the distribution
for small $x$ (which contributes strongly to $I$), we obtain with this
method values that are slightly larger than that quoted above for the dilute
limit.

To summarize, the behavior found in a large range of \NB\ is in support of
recent treatments of the problem by means of renormalization group and
scaling  analyses\cite{Schaefer,Schoot,Cates88}
and in contrast to earlier claims\cite{Guj} that the Schulz distribution,
Eq.(\ref{eq:p}), will hold independent of the degree of overlap between the
chains.

\subsection{Specific Heat.}
\label{sec:CV}
\begin{figure}
\centerline{\epsfysize=6cm
\epsfig{file=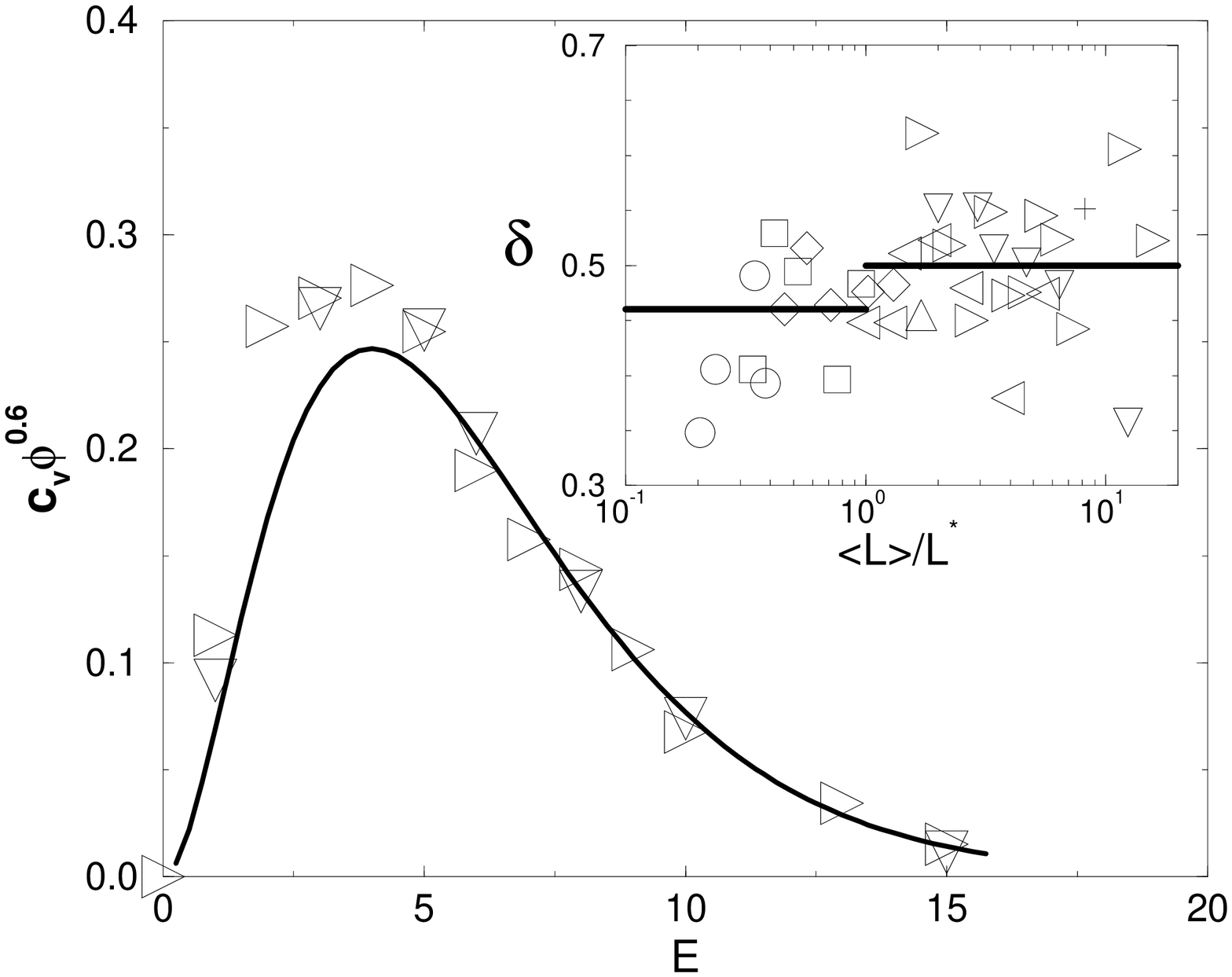,width=55mm,height=80mm,angle=-90}}
\vspace*{0.4cm}
{\small FIG.~7. 
The specific heat per monomer $\CV$.
We plot $\CV \phi^{\grow_s}$ versus scission energy $E$ which is
here equivalent to plot against the inverse temperature.
The line is the prediction in the semidilute regime.
Insert: Scaling-plot of $\CV \Lav/E^2=\delta$ versus number of blobs \NB\
in agreement with Eq.(\protect\ref{eq:CV}).
Symbols for the various densities as in Fig.~\protect\ref{fig:L}a.
\label{fig:CV}}
\end{figure}

A quantity that may be accessible experimentally is the specific heat \CV
(see also Tab.~\ref{tab:phiE}).
Due to its variable number of broken bonds (chain ends) EP should absorb or
release energy as the temperature, i.e. $E$, is varied. In Fig.~\ref{fig:CV}
we plot the specific heat per monomer $\CV
\phi^{\grow_s}$ versus scission energy $E$ for two densities within the
semidilute regime. We compare with the prediction of Eq.~(\ref{eq:CV}) where
we have used the exponents $\grow_s=0.6$ and $\delta_s=0.5$ and the
amplitude $A_s=4.4$ estimated in Sec.~\ref{sec:L}.
We find a qualitatively good agreement, especially, as espected,
for larger chains ($E > 5$).

In the insert we check the prediction $\CV \Lav/E^2 = \delta$
(plotted versus the numbers of blobs \NB) for all configurations with
$\Lav > 5$. Although the statistics is too poor to separate the extremely
small difference between $\delta_d$ and $\delta_s$ the data `collapse'
around $\delta=0.5 \pm 0.1$  is qualitatively satisfactory for this
ambitious test.

Note that although this heat capacity has some resemblence to that
predicted for stepwise living polymers in the present of
initiators, the latter systems are fundamentally different in that
they can show a (possibly rounded) phase transition at nonzero temperature
\cite{Pfeuty,Greer96}.  In the present system no critical phenomena are
observed at finite temperature; this is confirmed by the fact that the
measured heat capacity becomes system-size independent in the limit of large
systems (compare Tab.~\ref{tab:FS}).

\subsection{Conformational Properties.}
\label{sec:conf}
Following ref.\cite{Wolferl}
the average over all chains of the mean-square end-to-end distance \Ree\ and
the radius of gyration
\Rgg\  have been measured and are listed in Tab.~\ref{tab:phiE}.
To obtain meaningful results for chain size, when large chains are present,
we must however `unwrap' the  periodic box starting from the position of
one chain end, using the bond vectors between consequent monomers of the
chain. Not surprisingly the mean bond length \bb\ is nearly identical to the
ones obtained in the study by Paul et al.~\cite{Wolferl} and is likewise
slightly decreasing with density. Also we observe (for sufficiently long
chains) a persistence  length $\l_p \approx 1.1 b$.
We include in the table the mean interchain distance
$H=(8\Lav/\phi)^{1/3}$ which is of relevance in the context of diffusion
controlled scission and recombination events~\cite{WMCdynam}.

\begin{figure}
\centerline{\epsfysize=6cm
\epsfig{file=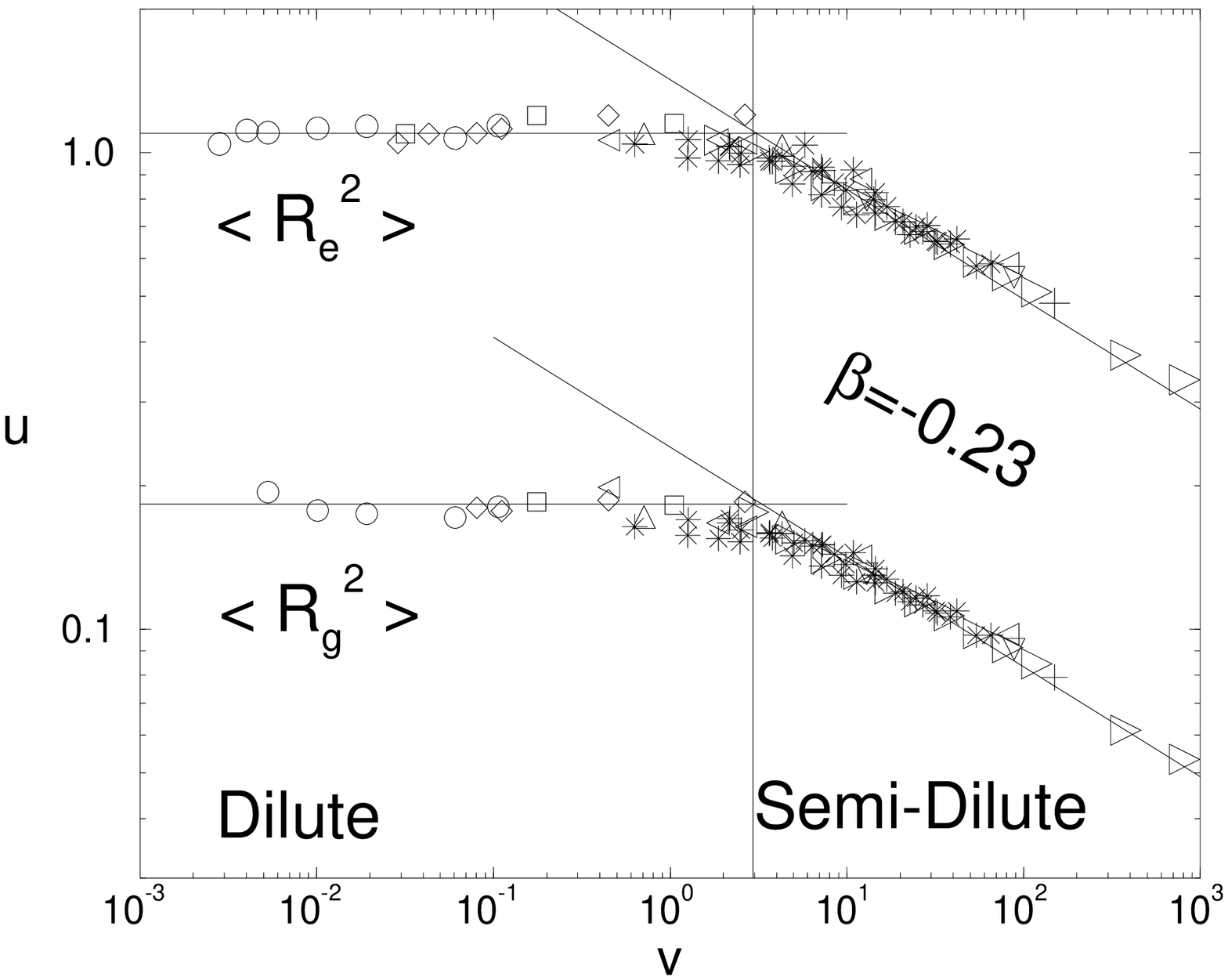,width=55mm,height=80mm,angle=-90}}
\vspace*{0.4cm}
{\small FIG.~8. 
Crossover scaling plot of the reduced mean-square end-to-end distance
$u_e=\Ree/\Rs^2$ and the reduced radius of gyration $u_g=\Rgg/\Rs^2$
vs the scaling variable $v = (\Rs/H)^3$. Same symbols as in
Fig.~\ref{fig:L}a, asterisks denote monodisperse chains.\label{fig:Rscal}}
\end{figure}

In Fig.~\ref{fig:Rscal} we demonstrate that the mean chain sizes for EP
follow the same universal function as conventional quenched polymers.
We want to compare sizes of chains of given mean-chain length with the
size of swollen dilute chains of the same length; therefore we define
$\Rs = b \Lav^{\nu}$.
We plot the reduced average chain size $u_e=\Ree/\Rs^2$ and $u_g=\Rgg/\Rs^2$
for the end-to-end distance and the radius of gyration respectively as
function of the scaling variable $v=(\Rs/H)^3$. This choice of variable,
rather than the alternative $\phi/\phistar$ (to which $v$ is proportional)
enables a comparison between EP  and conventional monodisperse polymers
(asterisks in Fig.~\ref{fig:Rscal}).
Data for the latter is taken from Paul et al~\cite{Wolferl}.
In the dilute regime we have $\Rend \approx
\Rgyr
\approx \Rs$ and the  scaling function $u$ approaches a constant as can be
seen in Fig.~\ref{fig:Rscal}.
Note that the plateau for $u_e$ is slightly above 1 in agreement with the
persistence length given above.
In the semi-dilute limit the chains are Gaussian on length scales larger
than the blob size $\xi$ (i.e. $\Rgyr \propto \Lav^{1/2}$)
implying the scaling $u \propto v^{-\beta}$
with exponent $\beta=(2\nu-1)/(3\nu - 1) \approx 0.23$.
It is remarkable that, on this scaling plot, EP and conventional monodisperse
polymers are  nearly indistinguishable;
the two universal functions are virtually identical to within numerical
accuracy.  Note the location of the crossover density at $v^* \approx 3 \pm
1$, i.e. $\Rend \approx \Rs \approx 1.4 H$.
A consistency check with the estimates of the amplitudes $P$ and $Q$ from
Eq.~(\ref{eq:PQ}) gives a slightly lower value $v^* = P Q^{3\nu-1} (b/a)^3
/ 8 \approx 1$.
However, in view of the error in locating the crossover values and the large
exponents involved in the estimation of $P$ and $Q$ this is
reasonably consistent.

\begin{figure}
\centerline{\epsfysize=6cm
\epsfig{file=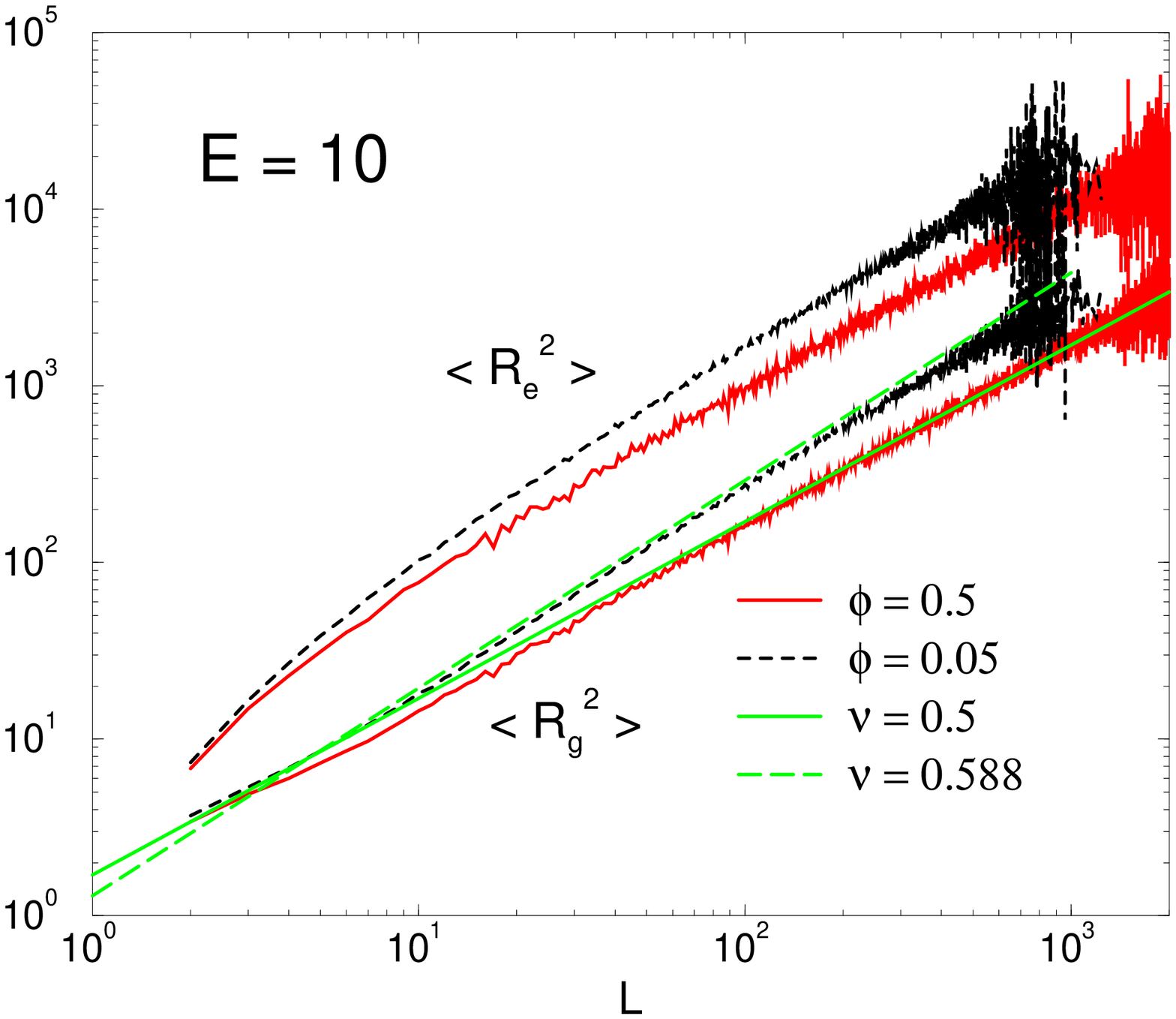,width=55mm,height=80mm,angle=-90}}
\centerline{\epsfysize=6cm
\epsfig{file=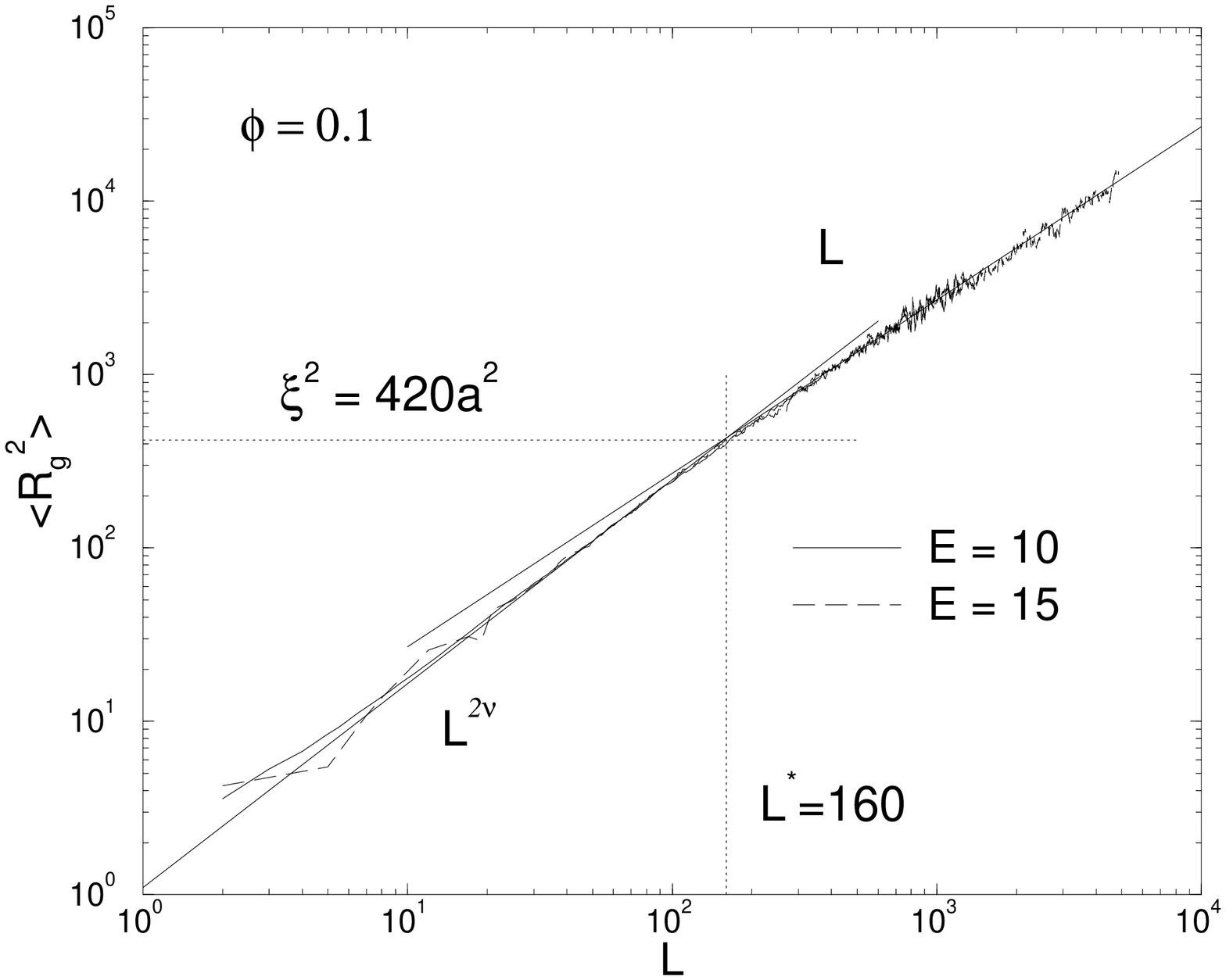,width=55mm,height=80mm,angle=-90}}
\vspace*{0.4cm}
{\small FIG.~9. 
Variation of chain size with chain length $L$.  (a) At very low density (e.g.
$\phi=0.05$ (dashed lines)) the chain are swollen $R \propto L^{\nu}$
with $\nu \approx 0.6$. At density $\phi=0.5$  one has a melt of Gaussian
chains ($\nu=0.5$).
(b) For a semi-dilute system, as for $\phi=0.1$ at $E=10$ and $E=15$,
one can identify two regimes. Up to sizes $\xi$ the chains are swollen,
on sizes larger than $\xi$ they are Gaussian.\label{fig:RL}}
\end{figure}

This conclusion is corroborated in Fig.~\ref{fig:RL} where we show the
distribution of chain sizes, $\Rgg_L$ and $\Ree_L$, averaged not over
all chains present (as considered above) but over all chains of length $L$,
plotted against $L$ (rather $\Lav$). For the melt density $\phi=0.5$ we
find
$\Ree_L \approx 6 \Rgg_L \propto L^{2\nu}$ with a Flory exponent
$\nu \approx 1/2$.
For the much smaller semi-dilute density $\phi=0.01$ we obtain
swollen chains with an excluded volume
exponent $\nu \approx 0.6$ (or perhaps slightly larger).

Fig.~\ref{fig:RL}b shows the distribution of
$\Rgg_L$ at an intermediate density which can be used to determine
the average size, $\xi$, of a blob containing $\Lstar$ monomers.
For $L \ll \Lstar$ and $\Rgg_L \ll \xi^2$, i.e.
within the blob, the excluded volume interactions dominate and
$\Rgg_L \propto L^{1.2}$.
For larger $L$ (i.e. also larger distances) the chains become
Gaussian $\Rgg_L \propto L$. We may measure the values of $\Lstar$ and $\xi$
directly from the crossover between both regimes and compare them with the
values obtained above. From the intersection point of
the two slopes in Fig.~\ref{fig:RL}b one has $\xi^2 \approx 420\;a^2$ and
$\Lstar \approx 160$.
Remembering that every monomer occupies 8 sites of the lattice
one obtains as a consistency check roughly $\phi = 8 \Lstar/\xi^3 \approx 0.13$
which nearly matches the actual density of the system.
Fig.~\ref{fig:RL}b confirms (within numerical accuracy)
the sharp crossover between asymptotic forms as discussed in connection with
Fig.~\ref{fig:L}b and is another of our main results.

\section{Finite-Size Effects.}
\label{sec:FS}

Since no chain can be larger than the total number of monomers present
in the system,
the exponential MWD must break down whenever the mean chain length becomes
too large,
i.e. when the average number of chains per box $\Mav \approx \Lmax/\Lav$ of
order one. (For the same reason it must always break down in the high
molecular weight tail, in any case.)
Note that for the highest energy $E=15$ used we find typically
an average chain number $\Mav \approx 10$.
In fact, the (rather noisy) MWD show even then qualitatively an
exponential decay.

\begin{figure}
\centerline{\epsfysize=6cm
\epsfig{file=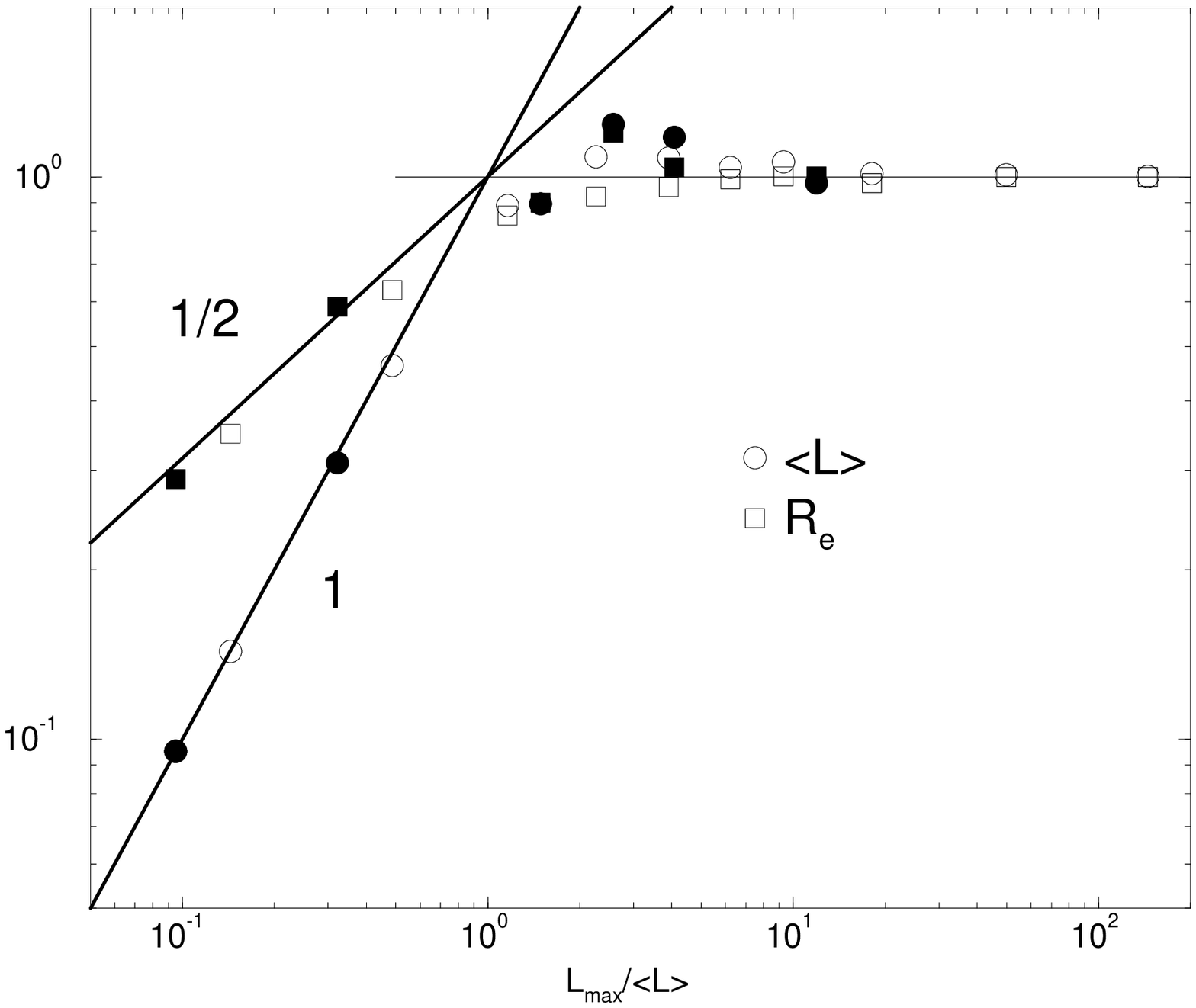,width=55mm,height=80mm,angle=-90}}
\centerline{\epsfysize=6cm
\epsfig{file=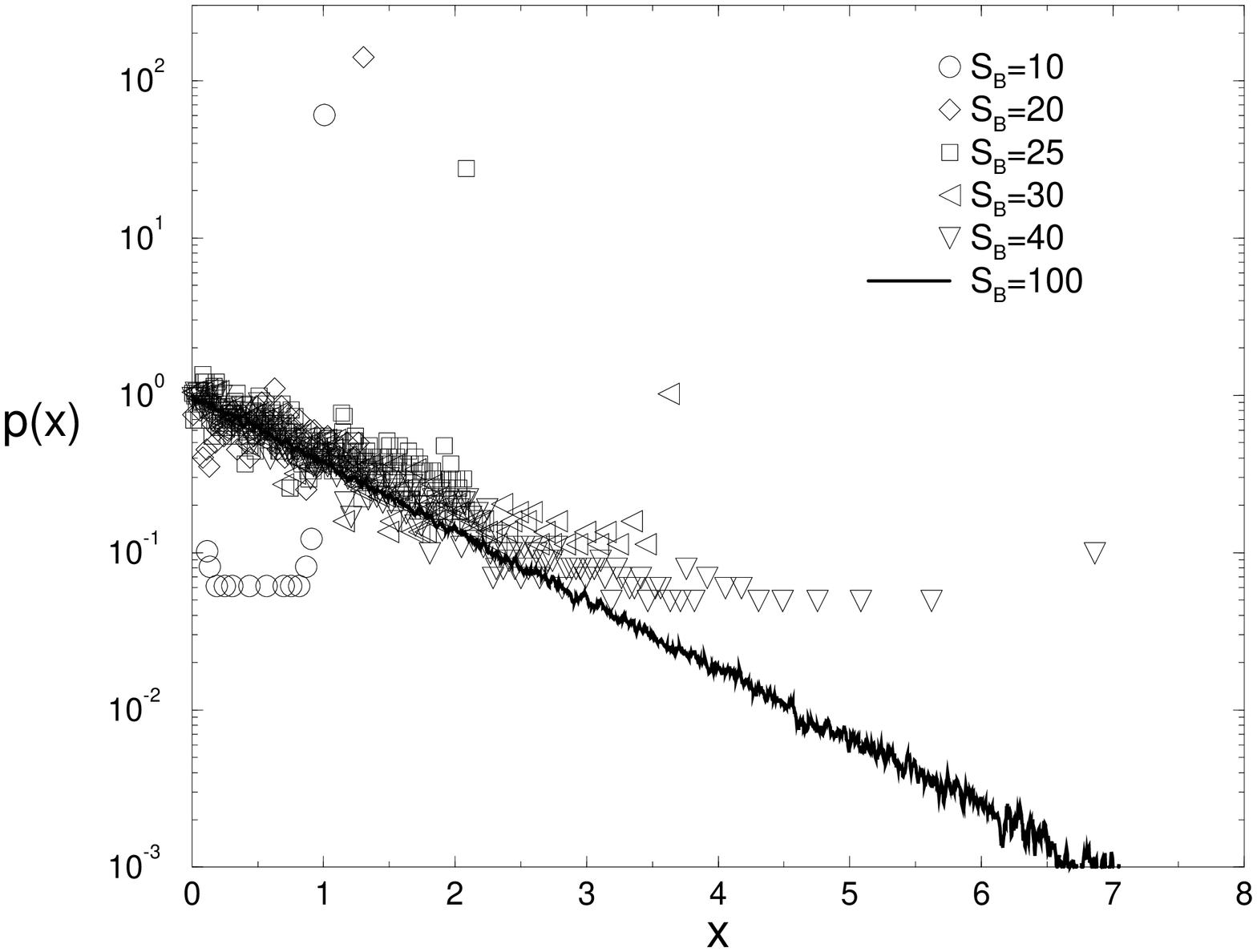,width=55mm,height=80mm,angle=-90}}
\vspace*{0.4cm}
{\small FIG.~10. 
Systematic variation of the system dimension \Lbox.
(a) Reduced chain length \Lav\ and end-to-end size \Rend\ 
versus $\Lmax$ reduced by the average chain length of an infinite system.
Solid symbols $E=15$, open symbols, $E=10$.
(b) MWD for $\phi=0.5$ at scission energy $E=10$ for various system sizes
as indicated in the figure.\label{fig:FS}}
\end{figure}

In order to understand finite-size effects quantitatively
we have explictly performed a systematic finite-size study for
$E=10$ and $E=15$ at density $\phi=0.5$ which is summarized in
Tab.~\ref{tab:FS}.
In Fig.~\ref{fig:FS}a the mean chain length \Lav\ and the end-to-end
distance \Rend\, reduced by their asymptotic
values for infinite systems
(taken from the predictions in Section~\ref{sec:Theory}
together with the amplitudes obtained above), are plotted against
an obvious finite-size scaling variable, which is
the number of monomers in the box $\Lmax$ divided by the average chain
length of an infinite system, \Lav.
We see that the systems for which we have presented results above are indeed
lying in the asymptotic plateau region.
(Additionally, this confirms the amplitudes and exponents presented already
above.) 
The scaling curves for the mean chain length and the specific heat
(not shown) are  -- not surprisingly --
similar, the specific heat being the noisier quantity.
Both decay in the small system limit with the system size (slope one),
implying the formation of a single long chain of length
$\approx \Lmax \propto \phi$.
Hence, the effective growth exponent \grow\ is likely to be overestimated
in computational studies on relatively small systems.
From $\Lmax/\Lav \propto \phi^{1-\grow}$ we see that increasing
the density at fixed scission energy should decrease the finite-size effects
shifting the system to the right in fig.~\ref{fig:FS}.
Note that at intermediate system sizes above $\Lmax/\Lav \approx 1$
the mean chain length is slightly {\em larger}
than the asymptotic value (see Tab.~\ref{tab:FS}).

For the chain size in the limit of strong finite size effects our results are
qualitatively consistent with a square root dependency, $R \propto
\Lmax^{1/2}$,  rather than a $1/3$ exponent for a single compact chain;
this is because the chain size is defined by `unwrapping' the
periodic box starting from one chain end (as mentioned above) and nothing
prevents a single Gaussian chain from wrapping repeatedly to fill the
periodic box. In Fig.~\ref{fig:FS}b we see directly from
the MWD how the transition to the single chain `phase' occurs. For small
systems ($\Lbox
\leq 30$) we find clear peaks in the distribution at $L=\Lmax$ which
disappear as we further increase the system size.
This finite-sizes study confirms unambiguously that
the configurations presented above are indeed large enough --
at least for the static properties.
It is however not so clear that this is still holds for dynamic properties
in the limit of large barrier $B$, and we will address this issue in
the second part of this work~\cite{WMCdynam}.

\section{Discussion.}
\label{sec:Discussion}

In the present Monte Carlo simulation of static (and dynamic\cite{WMCdynam})
bulk properties of EP we have used a greatly improved version of a recently
proposed  algorithm\cite{YA1}.
The original data structure based on the BFM polymer chain model was completely
altered, recognizing that the fundamental entities in EP systems are not
chains but monomers, or (in our algorithm) the bonds connecting these. This
makes the algorithm much more powerful, and allows us to study systems across
three decades in monomer density with up to
$75,000$ monomers per configuration (for $\phi=0.6$ on a DEC Alpha
workstation). With this algorithm we can equilibrate systems with mean-chain
length up to
$\Lav
\approx 5000$ in the melt, which is at least one orders of magnitude larger
than any other previous study\cite{Kro,YA1}. This was achieved with
negligible finite size effects as we have explictly checked.

Therefore, we are able to compare simulational results to data
extracted from laboratory  observations and to the scaling predictions of
theoretical treatments, known to hold in the asymptotic limit of
sufficiently long chains. Some earlier numerical observations which showed
discrepancies with analytic results could thereby be traced to finite-size
effects arising from small systems and/or short chains.

In the present investigation we simulated EP where the formation of rings is
not allowed. (A study for systems containing rings is
underway\cite{WSCMring}.) The nonbonded monomer interactions have been set
to zero (athermal conditions) so that the overall picture was not complicated
by phase separation at low temperature. For simplicity the chains
are modelled as totally flexible.

At variance to some recent computational studies\cite{Kro}
we find that the static properties in the asymptotic limit of large $\Lav$
agree well with analytical predictions of the scaling
theory\cite{Cates88,CatesCandau} and renormalization group studies
\cite{Schaefer,Schoot}:

\begin{enumerate}
\item
The mean chain length varies as
$\Lav = \Lstar \left(\phi/\phistar\right)^{\grow}
\propto \phi^{\grow} \exp(\delta E)$
with density $\phi$ and end-cap energy $E$.
As predicted analytically\cite{Schoot,Cates88},
we find the exponents $\grow_d=\delta_d=1/(1+\gamma)=0.46$ in the dilute regime
and $\grow_d=0.6,\delta_d=0.5$ in the semi-dilute limit.
We confirmed the scaling behavior of the mean number of blobs
$\NB=\Lav/\Lstar$ versus the reduced density $\phi/\phistar$. This shows
a rather sharp crossover which enables clean crossover lines $\phistar(E)$
and $\Lstar(E)$ have been located.
\item
In the dilute limit the MWD scales consistently
with a Schulz distribution $\cN(s) \propto s^{\gamma-1} \exp(-s)$
with a scaling variable $s = \gamma L/\Lav$.
In the semi-dilute regime for large enough \Lav\ the MWD decays exponentially
with chain-length $\cN(s) = \exp(-s)$ where the scaling variable
becomes the reduced chain length $s=L/\Lav$.
Between both limiting regimes we observe a relatively gradual crossover
at $\NB \approx 1$, the overlap threshold of the polymers.
The extremely large size of our chains, allied to a careful analysis of the
size distribution in the dilute limit, allows us to extract an accurate
estimate of the self-avoiding walk susceptibility exponent
$\gamma = 1.165 \pm 0.01$.
\item
A satisfactory scaling collapse of the specific heat in both density regimes
was obtained by plotting $\CV \Lav/E^2 = \delta \approx 0.5$ versus \NB.
As suggested earlier\cite{YA1}, the maximum of the specific heat per
monomer \CV\ occurs at $E=2/\delta=4$ for high concentrations $\phi$.
\item
Chain conformations are described within numerical accuracy by
the same universal functions as for conventional polymers.
\item
As in the case of conventional polymers, the chains are swollen within the
excluded volume blobs and Gaussian at larger distances - this has been made
directly evident from the scaling of coil size against
chain length within a single system.
\end{enumerate}

We believe that the present work unambiguously confirms the scaling results
for an idealized model of EP based on the classical behavior of conventional
quenched polymers \cite{Cates88}. Accordingly, it leaves completely
unanswered the question of how, in some experimental systems to which the
model appears closely applicable, a growth exponent $\grow \simeq 1.2$ is
convincingly argued to arise\cite{Schurtenberger}. This question remains
open, but clearly must involve physics not in the present model, which is
based on assuming a fixed scission energy, athermal excluded volume
interactions, and the absence of rings.

A parallel version of the algorithm has also been implemented and
used for some of the results presented above. It forms the basis of
the natural extension of this work to the question of
EP dynamics, which will be published elsewhere \cite{WMCdynam}.

\section*{Acknowledgement}
The authors are indebted to
J.~P.~Desplat, Y.~Rouault, M.~M\"uller, P.~van~der~Schoot and F.~Lequeux
for valuable discussions and assistance during the present investigation.
JPW acknowledges support by EPSRC under Grant GR/K56223 and is
indebted to D. P. Landau for hospitality in the Center for Simulational
Physics at the University of Georgia.
AM acknowledges the hospitality of the EPCC in Edinburgh (TRACS program),
and the support by the {\em Deutsche Forschungsgemeinschaft} (DFG)
under grant No. 436-BUL 113/45, and grant No. 301/93 given by the
{\em Bulgarian Ministry for Science and Education}.


\end{multicols}


\begin{table}[t]
\begin{tabular}{|l|c||c|c|c|c|c|c|} \hline
$\phi$ &  E &\Mav & \Lav& \CV & \Rend&\Rgyr&$H$\\ \hline
0.001
       & 13 & 16  & 62    & 0.7  & 32   & 13   & 79     \\
       & 15 & 8.5 & 124   & 0.7  & 50   & 20   & 98    \\ \hline
0.005
       & 13 & 41  & 125   & 0.5  & 51   & 2    & 56    \\
       & 15 & 17  & 311   & 0.3  & 86   & 34   & 78    \\ \hline
0.01
       & 13 & 60  & 168   & 0.5  & 61   & 24   & 51    \\
       & 15 & 24  & 424   & 0.26 & 105  & 41   & 69    \\ \hline
0.05   & 10 & 498 & 101   & 0.45 & 42   & 17   & 25    \\
       & 15 & 46  & 1110  & 0.13 & 165  & 64   & 56    \\ \hline
0.1
       & 8  & 218 & 57    & 0.58 & 29   & 12   & 17    \\
       & 10 & 81  & 155   & 0.31 & 50   & 20   & 23    \\
       & 13 & 20  & 641   & 0.1  & 112  & 44   & 37     \\
       & 15 & 7.5 & 1777  & 0.06 & 163  & 68   & 51    \\ \hline
0.4
       & 6  & 970 & 52    & 0.36 & 23   & 10   & 10    \\
       & 8  & 360 & 139   & 0.23 & 39   & 16   & 14     \\
       & 10 & 133 & 375   & 0.128& 65   & 26   & 20    \\
       & 13 & 30  & 1898  & 0.05 & 132  & 54   & 33   \\
       & 15 & 13  & 4017  & 0.02 & 196  & 81   & 43    \\ \hline
0.5
       & 6  & 1058& 59    & 0.29 & 24   & 10   & 10   \\
       & 7  & 642 & 98    & 0.24 & 31   & 13   & 12  \\
       & 8  & 391 & 160   & 0.22 & 40   & 16   & 14  \\
       & 9  & 237 & 263   & 0.16 & 52   & 21   & 16  \\
       & 10 & 145 & 432   & 0.1  & 67   & 27   & 19    \\
       & 13 & 32  & 1968  & 0.05 & 140  & 58   & 31  \\
       & 15 & 13  & 5113  & 0.02 & 231  & 92   & 43    \\ \hline
0.6    & 10 & 157 & 480   & 0.11 & 68   & 28   & 19  \\
\hline
\end{tabular}
\vspace*{0.5cm}
\caption{Summary of measured static quantities for configurations
with $\Lav > 50$.
Three decades of volume fractions between $\phi=0.001$ and $\phi=0.6$
have been sampled with bond energies up to $E = 15$.
Note that $\Lbox(\phi < 0.1) = 200 a$ and $L_{Box}(\phi \ge 0.1) = 100 a$.
Quantities tabulated:
the mean number of chains \Mav,
the mean chain length $\Lav$, the specific heat per monomer \CV,
the mean end-to-end distance $\Rend$,
the mean gyration radius $\Rgyr$ and
the average distance between chains $H$.
All length scales are given in units of the lattice constant $a$.
}
\label{tab:phiE}
\end{table}

\begin{table}[t]
\begin{tabular}{|l|c||c|c|c|c|c|} \hline
E   &\Lbox/\Lmax & \Mav & \Lav & \CV  &$\Rend$&$\Rgyr$\\ \hline
10  & 10/62    & 1.0  & 62  & 0.01 &  23 & 10     \\
    & 15/210   & 1.1  & 200 & 0.05 &  42 & 17      \\
    & 20/500   & 1.5  & 384 & 0.09 &  57 & 23      \\
    & 25/976   & 2.6  & 468 & 0.12 &  62 & 25      \\
    & 30/1687  & 4.1  & 466 & 0.11  & 64 & 26      \\
    & 35/2679  & 6.4  & 448 & 0.11  & 66 & 27      \\
    & 40/4000  & 9.3  & 458 & 0.13 &  67 & 27      \\
    & 50/7812  & 18   & 436 & 0.10 &  65 & 27     \\
    & 70/21437 & 50   & 435 & 0.11 &  67 & 27     \\
    & 100/62500& 145  & 432 & 0.10 &  67 & 27 \\  \hline
15  & 20/500   & 1.0  & 499 & 0.001& 67  & 27\\
    & 30/1687  & 1.1  & 1627 & 0.01& 136 & 55 \\
    & 50/7812  & 2.0  & 4706 & 0.02 & 208 & 80 \\
    & 60/13500 & 2.5  & 6512 & 0.02 & 277 & 104 \\
    & 70/21437 & 4.4  & 6180 & 0.03  & 240 & 95 \\
    & 100/62500& 12.7 & 5113 & 0.02 & 231 & 92 \\ \hline
\end{tabular}
\vspace*{0.5cm}
\caption{Variation of system size \Lbox\ for $\phi=0.5$ for two
high scission energies.
We give the average number of chains \Mav,
the mean chain length \Lav,
the specific heat \CV,
the end-to-end distance \Rend\
and the radius of gyration \Rgyr.
}
\label{tab:FS}
\end{table}


\begin{references}

\bibitem{CatesCandau}
M.~E.~Cates and S.~J.~Candau.
J.~Phys.~Condens.~Mattter {\bf 2}, 6869 (1990).

\bibitem{Sulfur}
R. L. Scott, J. Phys. Chem. {\bf 69}, 261 (1965).

\bibitem{Pfeuty}
J.~C.~Wheeler, S.~J.~Kennedy, and P.Pfeuty, 
Phys. Rev. Lett. {\bf 45}, 1748 (1980);
S.~J.~Kennedy and J.~C.~Wheeler, J.Phys. Chem. {\bf 78}, 953 (1984).

\bibitem{Selen}
G.~Faivre and J.~L.~Gardissat, Macromolecules {\bf 19}, 1988 (1986).

\bibitem{Proteins}
F.~Oozawa and S.~Asakura, {\em Thermodynamics in the Polymerization
of Proteins} (Academic Press, New York, 1975).

\bibitem{Szwarc56}
M.~Szwarc, {\em Nature}, {\bf 178}, 1168 (1956).

\bibitem{Greer96}
LP polymerize (above or below some critical temperature)
starting from an imposed density of initiators by reversible addition
of monomers on active chain ends. They are held together by (strong) covalent
carbon-to-carbon bonds, hence, they do not break in the middle of
the polymer chain. Neither do LP combine together to make larger LP or
to form rings. For a recent review see
S.~C.~Greer, Advances in Chemical Physics {\bf 96} 261 (1996).

\bibitem{Clausen}
T. M. Clausen, P. K. Vinson, J. R. Minter, H. T. Davis, Y. Talmon, and
W. G. Miller, J. Phys. Chem. {\bf 96}, 474 (1992).

\bibitem{likepolymer}
R. Messager, A. Ott. D, Chatenay, W. Urbach, D.Langevin,
Phys. Rev. E, 60 (1988) 1410;
J.~Appel and  G.~Porte, Europhysics Lett. {\bf 12}, 185 (1990).

\bibitem{Cates88}
M.~E.~Cates, {\em J. de Physique} {\bf 49} 1593 (1988).

\bibitem{Faet}
E. Faetibold and G. Waton, Langmuir {\bf 11}, 1972 (1995).

\bibitem{Bouchaud}
A. Ott, J. P. Bouchaud, D. Langevin and W. Urbach, 
Phys. Rev. Lett. {\bf 65}, 2201 (1990);
J. P. Bouchaud, A. Ott, D. Langevin and W. Urbach, 
J. Phys. II (France), {\bf 1}, 1465 (1991).

\bibitem{OShaug}
B. O'Shaughnessy and J. Yu, Phys. Rev. Lett. {\bf 74}, 4329 (1995).

\bibitem{Flory53}
P.~J.~Flory, {\em Principles of Polymer Chemistry}
(Cornell University Press, Ithaca, NY, 1953).

\bibitem{Schaefer}
L. Sch\"{a}fer, Phys. Rev. {\bf B 46}, 6061 (1992).

\bibitem{Schoot}
P. van der Schoot, Europhys. Lett. {\bf 39} 25 (1997).

\bibitem{Berret}
J. F. Berret, J. Appell and G. Porte, Langmuir {\bf 9}, 2851 (1993).

\bibitem{Schurtenberger}
G.~Jerke, J.S.~Pedersen, S.U.~Egelhaaf, P.~Schurtenberger,
Physical Review E, {\bf 56} 5772 (1997);
P. Schurtenberger, C. Cavaco, F. Tiberg and O. Regev, 
Langmuir {\bf 12}, 2894 (1996).

\bibitem{MilchevPotts}
A.~Milchev, Polymer {\bf 34}, 362 (1993);
A.~Milchev and D.~P.~Landau,  Phys. Rev. E {\bf 52}, 6431 (1995).
A.~Milchev and D.~P.~Landau, J. Chem. Phys. {\bf 104}, 9161 (1996).

\bibitem{YA1}
Y.~Rouault, A.~Milchev, Phys. Rev. E {\bf 51}, 5905 (1995).
\bibitem{YA2}
Y.~Rouault, A.~Milchev, J. Phys. II France {\bf 5}, 343 (1995).
\bibitem{YA3}
Y. Rouault and A. Milchev, Phys. Rev. E {\bf 55}, 2020 (1997).

\bibitem{Kro}
M. Kr\"{o}ger and R. Makhloufi, Phys. Rev. E {\bf 53}, 2531 (1996);
W.~Carl, R.~Makhloufi, M.~Kroger, Journal de Physique II {\bf 7}, 931 (1997).


\bibitem{BFM}
I.~Carmesin and K.~Kremer, Macromolecules, {\bf 21}, 2819 (1988).

\bibitem{Wolferl}
W.~Paul, K.~Binder, D.~Herrmann, and K.~Kremer,
J. Chem. Phys. {\bf 95}, 7726 (1991).

\bibitem{Marcus}
M.~M\"uller, K.~Binder, Computer Physics Communications, {\bf 84} 173 (1994);
M.~M\"uller, K.~Binder, Macromolecules {\bf 28}, 1825 (1995).

\bibitem{WPB1}
J.~Wittmer, W.~Paul, K.~Binder, Macromolecules, {\bf 25}, 7211 (1992).

\bibitem{ringeffect}
R. Petschek, P. Pfeuty and J. C. Wheeler, Phys. Rev. A 34
(1986) 2391. See also, M. E. Cates, J. Physique Lett. {\bf 46}, L837 (1985).

\bibitem{WMC1}
J.~P.~Wittmer, A.~Milchev, and M.~E.~Cates,
Europhysics Letters, {\bf 41}, 291 (1998).

\bibitem{WSCMring} J.~P.~Wittmer, P.~van~der~Schoot, A.~Milchev, M.~E.~Cates,
in preparation.

\bibitem{branchmic} F. Lequeux and S.J. Candau, in {\it Structure and Flow
in Surfactant Solutions},  C. A. Herb and R. K. Prud'homme, Eds., ACS Books,
Washington (Symposium Series 578) (1994)

\bibitem{WMCdynam}
J.~P.~Wittmer, A.~Milchev, and M.~E.~Cates, in preparation.

\bibitem{Binderbook79}
See e.g., {\em Monte Carlo Methods in Statistical Physics},
edited by K.~Binder (Springer Verlag, Berlin 1979).

\bibitem{Guj}
P. D. Gujrati, Phys. Rev. B {\bf 40}, 5140 (1989).

\bibitem{deGennes79}
P.~G.~de~Gennes, in {\em Scaling Concepts in Polymer Physics}
(Cornell University Press, Ithaca, NY, 1979), Chap. 1.

\bibitem{AndreyOfflattice}
I. Gerroff, A. Milchev, K. Binder, and W. Paul, J. Chem. Phys. {\bf 98}, 6526 (1993); 
A. Milchev, W. Paul, and K. Binder, J. Chem. Phys. {\bf 99}, 4786 (1993).

\bibitem{shortbonds}
This is rather unfortunate the relaxation times being much faster using
the full $108$ bonds available and no difference having been observed
in the static properties where the (extremely unlikely) crossed bonds do
not contribute. However, the rare event of fixed monomers (in the limit
of high barrier $B$ where crossed bonds cannot be broken sufficiently fast)
changes completely the dynamic properties from Rouse to Reptation behaviour
(due to the view artifical fixed obstacles)
as obvious from the mean-square displacement.\cite{WMCdynam}
We are indebted to Y.~Rouault and M.~M\"uller for pointing our attention
on this technical difficulty related to the lattice nature of the
underlying BFM.

\bibitem{CatesGranek} R. Granek and M. E. Cates, 
J. Chem. Phys, {\it 96}, 4758--4767 (1992).

\bibitem{desCloizJannink}
J.~des Cloizeaux and G.~Jannink, Polymers in Solution, Clarendon Oxford (1990).

\end{references}
\end{document}